\documentclass[aip,jcp,reprint,twocolumn,showpacs,nofootinbib]{revtex4}
\usepackage[utf8x]{inputenc}

\usepackage{microtype}
\usepackage{enumitem}
\usepackage{amsmath,amssymb,bm}
\usepackage{graphicx}
\usepackage[toc,page]{appendix}
\usepackage{mathtools}

\newcommand{\pder}[2]{\frac{\partial #1}{\partial #2}}
\newcommand{\grad}[2]{\partial_{#2} #1 }

\newcommand{\stder}[2]{\frac{\d^2 #1}{{\d #2}^2}}
\newcommand{\mpder}[3]{\dfrac{\partial^2 #1}{{\partial #2}{\partial #3}}}

\newcommand{\vX}{\left|\bm{X}\right\rangle}

\newcommand{\vu}{\left|\bm{u}\right\rangle}
\newcommand{\vuo}{\left|\bm{u}^0\right\rangle}
\newcommand{\vut}{\left\langle\bm{u}\right|}
\newcommand{\vfnrm}{\left|{\bm{f}}^u\right\rangle}
\newcommand{\vf}{\left|\bm{f}\right\rangle}
\newcommand{\vfo}{\left|\bm{f}^0\right\rangle}

\newcommand{\scal}[2]{\left\langle{#1} \middle| {#2}\right\rangle}
\newcommand{\vv}[1]{\left|{\bm{v}_{#1}}\right\rangle}

\newcommand{\vR}{\bm{R}}
\newcommand{\vq}{\bm{q}}
\newcommand{\vm} {{\bm{m}}}
\renewcommand{\vr}{{\bm{r}}}

\newcommand{\dirx}{{\widehat{\bm{x}}}}
\newcommand{\diry}{{\widehat{\bm{y}}}}
\newcommand{\dirz}{{\widehat{\bm{z}}}}
\newcommand{\dirvm}{{\widehat{\bm{m}}}}
\newcommand{\hess}{{\mathcal{H}}}

\newcommand{\pbc}{{\tt pbc}}
\newcommand{\obc}{{\tt obc}}
\newcommand{\fcc}{{\tt fcc}}

\newcommand{\bd}{{\tt BD}}
\newcommand{\lr}{{\tt LR}}

\newcommand{\unitl}{{l_0}}

\newcommand{\npcube}{3375}
\newcommand{\npfcc}{3430}
\newcommand{\npdisord}{1688}
\newcommand{\ndissamples}{80}
\newcommand{\mlow}{0.056}
\newcommand{\mhigh}{0.064}

\renewcommand{\d}{\textrm{d}}
\newcommand{\e}{\textrm{e}}
\newcommand{\imu}{\textrm{i}}

\newcommand {\rij} {\bm{r}_{ij}}

\newcommand{\sumtwolines}[3]{\sum_{\substack{#1{}\\#2}}^{#3}}

\usepackage{ulem}
\usepackage{color}
\newcommand{\GP}[1]{\textcolor{blue}{#1}}
\renewcommand{\GP}[1]{{#1}}

\begin{document}

\title{
 Dynamic Elastic Moduli in Magnetic Gels: Normal Modes and Linear Response
}

\author{Giorgio Pessot}
\email{giorgpess@thphy.uni-duesseldorf.de}
\affiliation{Institut f\"ur Theoretische Physik II: Weiche Materie,
 Heinrich-Heine-Universit\"at D\"usseldorf, D-40225 D\"usseldorf, Germany}

\author{Hartmut L\"owen}
\email{hlowen@thphy.uni-duesseldorf.de}
\affiliation{Institut f\"ur Theoretische Physik II: Weiche Materie,
 Heinrich-Heine-Universit\"at D\"usseldorf, D-40225 D\"usseldorf, Germany}

\author{Andreas M.\ Menzel}
\email{menzel@thphy.uni-duesseldorf.de}
\affiliation{Institut f\"ur Theoretische Physik II: Weiche Materie,
 Heinrich-Heine-Universit\"at D\"usseldorf, D-40225 D\"usseldorf, Germany}

\date{\today}

\begin{abstract}
In the perspective of developing smart hybrid materials \GP{with} customized features, ferrogels and \GP{magnetorheological} elastomers allow a synergy of elasticity and magnetism.
The interplay between elastic and magnetic properties gives rise to a unique reversible control of the material behavior by applying an external magnetic field.
Albeit few works have been performed on the time-dependent properties so far, understanding the dynamic behavior is the key to model many practical situations, e.g.\ applications as vibration absorbers.
Here we present a way to calculate the frequency-dependent elastic moduli based on the decomposition of the linear response to an external stress in normal modes.
We use a minimal three-dimensional dipole-spring model to theoretically describe the magnetic and elastic interactions on the mesoscopic level.
Specifically, the magnetic particles carry permanent magnetic dipole moments and are spatially arranged in a prescribed way, before they are linked by elastic springs.
An external magnetic field aligns the magnetic moments.
On the one hand, we study regular lattice-like particle arrangements to compare with previous results in the literature.
On the other hand, we calculate the dynamic elastic moduli for irregular, more realistic particle distributions.
Our approach measures the tunability of the linear dynamic response as a function of the particle arrangement, the system orientation with respect to the external magnetic field, as well as the magnitude of the magnetic interaction between the particles.
The strength of the present approach is that it explicitly connects the relaxational modes of the system with the rheological properties as well as with the internal rearrangement of the particles in the sample, providing new insight into the dynamics of these remarkable materials.
\end{abstract}

\pacs{82.35.Np, 63.50.-x, 62.20.de, 75.80.+q}



\maketitle

\section{Introduction}\label{introduction}
The class of smart hybrid materials encompassing ferrogels and \GP{magnetorheological} elastomers stands out for their unique capability of combining magnetic properties with huge elastic deformability \cite{filipcsei2007magnetic,menzel2015tuned,lopez2016mechanics,odenbach2016microstructure}.
They typically consist of a permanently crosslinked polymer matrix in which magnetic colloidal particles are embedded.
The matrix is responsible for the elastic behavior typical of rubbers, while the particles magnetically interact with each other and with external magnetic fields.
These materials distinguish themselves by the fascinating ability of reversible on-demand tunability of shape and stiffness under the influence of external magnetic fields \cite{filipcsei2007magnetic,menzel2015tuned,odenbach2016microstructure,jarkova2003hydrodynamics,zrinyi1996deformation,stepanov2007effect,guan2008magnetostrictive,bose2009magnetorheological, gong2012full,evans2012highly,borin2013tuning} similarly to the tunability of viscosity in ferrofluids \cite{rosensweig1985ferrohydrodynamics,odenbach2003ferrofluids,odenbach2003magnetoviscous,odenbach2004recent,huke2004magnetic,ilg2005structure,holm2005structure,klapp2005dipolar,cattes2015condensation,peroukidis2015spontaneous,Klapp2016collective}.
This makes them ideal candidates for applications as soft actuators \cite{zimmermann2006modelling}, vibration absorbers \cite{deng2006development,sun2008study}, magnetic field detectors \cite{szabo1998shape,ramanujan2006mechanical}, and even as model systems to study aspects of hyperthermal cancer treatment \cite{babincova2001superparamagnetic,lao2004magnetic}.

The core feature of these materials is their magneto-mechanical coupling \cite{frickel2011magneto,allahyarov2014magnetomechanical,kaestner2013higher}, i.e.\ the way magnetic effects such as the response to an external magnetic field couple to the overall mechanical properties (e.g.\ strain or elastic moduli) and vice versa.
As was recently shown, such coupling is responsible for surprising properties such as superelasticity \cite{cremer2015tailoring}, a characteristic buckling of chains of particles under a perpendicular external magnetic field \cite{huang2015buckling}, qualitative reversal of the strain response \cite{allahyarov2014magnetomechanical}, volume changes due to mesoscopic wrapping effects \cite{weeber2015ferrogels2}\GP{, or tunability of the electrical resistance \cite{mietta2016anisotropic}}.
There are several key factors that can influence the magneto-mechanical coupling: the magnetic particle concentration \cite{varga2006magnetic,haider2015exceptionally,filipcsei2007magnetic}, the stiffness of the gel \cite{wisotzki2014tailoring}, or whether the magnetic moments of the particles can freely reorient or must instead rotate synchronously with the whole particle \cite{menzel2015tuned,weeber2012deformation}.
The particles can be chemically bound to the polymer network \cite{roeder2015magnetic,messing2011cobalt,frickel2011magneto} or be confined inside pockets of the matrix \cite{gundermann2014investigation,landers2015particle}.
Moreover, the magnetic material itself can either be ferro- \cite{messing2011cobalt} or (super)paramagnetic \cite{garcia2003mesoporous}.

Because of the variety of factors and parameters that can characterize ferrogels and magnetic elastomers, it is no surprise that they are receiving increasing attention from the modeling side.
In fact, gaining insight into the mechanisms underlying the magneto-mechanical coupling can be the key to devise smarter and more efficient materials.
Macroscopic theories rely on a continuum-mechanical description of both the polymeric matrix and the magnetic component \cite{jarkova2003hydrodynamics,brand2015macroscopic,brand2014macroscopic,allahyarov2014magnetomechanical,szabo1998shape,zubarev2012theory,bohlius2004macroscopic}, whereas mesoscopic approaches can take into account the granularity and discreteness of the magnetic particles \cite{cremer2015tailoring,ivaneyko2012effects,dudek2007molecular,wood2011modeling}.
On this mesoscopic level, simplified dipole-spring models represent a convenient approach to address effects originating on the magnetic particle level.
More precisely, in such models the particles carry a dipole magnetic moment and are linked with each other by a network of elastic springs.
Additionally, steric repulsion and other effects like orientational memory terms can be included
 \cite{annunziata2013hardening, pessot2014structural, tarama2014tunable, sanchez2013effects,cerda2013phase,ivaneyko2015dynamic}.
Finite-element descriptions are likewise employed to address mesoscopic particle-based effects \cite{spieler2013xfem,kaestner2013higher,raikher2008shape,stolbov2011modelling,gong2012full,han2013field}, and some works even resolve the individual polymers on the microscopic scale \cite{weeber2012deformation,weeber2015ferrogels}.
Moreover, in a coarse-graining perspective, some routes have been outlined to connect the different length scales listed above \cite{menzel2014bridging,pessot2015towards}.

Often in material science, one aims at determining the material parameters that characterize the system.
Fundamental quantities to describe the time-dependent mechanical behavior are the dynamic elastic moduli.
They, for instance, contain the information on the frequency-dependent stress response to imposed time-periodic deformations.
In the case of ferrogels and magnetic elastomers only few theoretical studies have so far addressed the dynamic properties in special cases \cite{tarama2014tunable,ivaneyko2015dynamic,belyaeva2016transient}.
In the present work we aim at calculating the dynamic (i.e.\ frequency-dependent) elastic moduli of ferrogels.
We use a minimal three-dimensional ($3$D) dipole-spring model with short-ranged steric repulsion between the magnetic particles.
Moreover, we consider the system around its equilibrium state of minimum total energy.
Overdamped motion of the particles is assumed, which is in general a reasonable assumption for colloidal polymeric systems.
\GP{We focus on regular and more disordered particle arrangements of finite size with open boundary conditions (\obc).
In our particle-based approach this simply refers to a detached finite assembly of particles.
This system is bounded in all three directions of space, in contrast to periodic boundary conditions (\pbc).}
We describe a semi-analytical approach using a simple, direct connection between the normal modes of the system and the linear response to an oscillating external stress.

The paper is structured as follows.
First, in section \ref{dip_spring_model} we present our minimal dipole-spring model including steric repulsion.
To find the equilibrium configurations under magnetic interactions, we use the methods as described in section \ref{equil_state}.
Then, in section \ref{norm_modes}, we determine the normal modes and in section \ref{static_elm_norm_modes} we connect them to the static linear elastic response of the system.
After that, in section \ref{dynamics}, we address the dynamic behavior of our system and show how to decompose it into the normal modes.
In section \ref{dynamic_elm}, we extend the elastic moduli expressions obtained in section \ref{static_elm_norm_modes} to the dynamical case and show the corresponding numerical results in sections \ref{cubic_result},  \ref{fcc_result}, and \ref{disord_result} before drawing our final conclusions in section \ref{conclusions}.
Appendix \ref{steric_param_app} lists the specific expressions used in modeling the steric repulsion, whereas appendices \ref{der_app} and \ref{hess_app} list in detail the employed expressions for gradients and Hessian matrices.
Appendix \ref{torq_app} describes in detail our procedure of obtaining a torque-free force field.
In appendix \ref{anyoung_app} we analytically estimate the Young moduli of regular lattices for comparison with our numerical results.
Last, in appendix \ref{loss_app} we present further data on the loss components of the dynamic moduli, supporting our results in the main text.

\section{Dipole-Spring Model}\label{dip_spring_model}
For simplicity we here work with a minimal $3$D dipole-spring model.
On the one hand, as a first approximation, we represent the magnetic moments by permanent point dipoles of constant magnitude.
Possible magnetic contributions due to the finite extension of the magnetic particles are not considered.
This is a valid approach for interparticle distances larger than the particle size (i.e.\ at low densities) \cite{biller2014modeling}.
In a simplified manner, spatial variations in dipole orientations and magnitudes due to their mutual feedback could be included in a subsequent step, see Ref.~\cite{allahyarov2015simulation}.
On the other hand, the interaction between the mesoscopic particles mediated by the polymeric matrix is, in general, non-linear \cite{pessot2015towards}.
However, since we are mainly interested in the linear elastic moduli for small displacements around the equilibrium positions of the particles, we confine ourselves to harmonic interactions in the present study.

Our system is made of $N$ identical spherical magnetic particles with positions $\vR_i = (R^x_i, R^y_i, R^z_i)$, $i =1 \dots N$.
To model the overdamped dynamics of the system, we consider viscous drag forces $-c \dot{\vR}_i$ during particle displacements, where the dot indicates the time derivative.
Each particle carries an identical magnetic dipole moment $\vm$ of magnitude $m=|\vm|$.
This situation reflects, for instance, the case of ferromagnetic or superparamagnetic particles under strong external magnetic fields.
Neighboring particles $i$ and $j$ are coupled by harmonic springs attached to the particle centers for simplicity.
The unstrained spring length $\ell_{ij}^0$ is set in the initial ground state particle configuration in the absence of any magnetic interactions, while the spring constants are given by $k/\ell_{ij}^0$.
Thus, $k$ is related to the overall elastic modulus of the system and long springs are weakened when compared to short ones.
We assume the polymeric matrix---here represented by the network of springs---to have vanishing magnetic susceptibility and therefore not to directly interact with magnetic fields.
If magnetic particles come too close to each other, they interact sterically.

The total energy $U$ of the system is the sum of elastic $U^{el}$, steric $U^{s}$, and magnetic $U^m$ energies \cite{pessot2014structural,annunziata2013hardening,cerda2013phase,sanchez2013effects}.
Elastic interactions are given by
\begin{equation}\label{eel}
 U^{el}= \frac{1}{2}\sum_{i \neq j} \frac{k_{ij}}{2} {\left( r_{ij} -\ell_{ij}^0 \right)}^2,
\end{equation}
where the sum runs over all particles $i$ and $j\neq i$.
Moreover, $k_{ij} = k/\ell_{ij}^0$ if particles $i$ and $j$ are connected by a spring and vanishes otherwise.
Furthermore, $\bm{r}_{ij}=\vR_j -\vR_i$ and $r_{ij}=|\bm{r}_{ij}|$.

We model the steric interactions using a repulsive potential inspired by the Weeks-Chandler-Andersen form \cite{weeks1971role} but with different exponents.
For instance, possibly absorbed polymer chains on the surfaces of the particles \cite{huang2015buckling} could result in a softer repulsion.
Our steric potential reads
\begin{equation}\label{est}
 U^{s}=\frac{1}{2}\sum_{i \neq j} v^{s} (r_{ij}),
\end{equation}
where
\begin{align}
v^{s} (r)&=\varepsilon^{s} \Bigl[ \left( {\frac{r}{\sigma^{s}}} \right)^{-4}- \left( {\frac{r}{\sigma^{s}}} \right)^{-2} - \left( {\frac{r_c}{\sigma^{s}}} \right)^{-4} +\left( {\frac{r_c}{\sigma^{s}}} \right)^{-2} \nonumber  \\
& \qquad \qquad +c^{s}\frac{{(r-r_c)}^2}{2} \Bigr]
\end{align}
for $r \le r_c$ and zero otherwise.
Here, $\varepsilon^{s}$ sets the strength of the steric repulsion, $\sigma^s$ characterizes the range of steric repulsion, and $r_c=\sigma^{s} 2^{1/2}$ is a cutoff distance.
The parameter $c^s$ is chosen such that altogether we have $v^{s} (r_c)=0$, ${v ^{s}}' (r_c)=0$, and ${v^{s}}'' (r_c)=0$ (see Appendix \ref{steric_param_app}).

Finally, the magnetic energy is given by the dipole--dipole interaction
\begin{equation}\label{emagn}
 U^{m}= \frac{\mu_0 m^2}{4\pi} \ \frac12 \sum_{i \neq j} \frac{r_{ij}^2 -3{(\dirvm\cdot\bm{r}_{ij})}^2}{r_{ij}^5},
\end{equation}
where $\dirvm=\vm/m$ and $\mu_0$ is the magnetic permeability of vacuum.
In the present work, we use reduced units as follows: lengths are given in multiples of $\unitl$, energies in multiples of $k \unitl$.
The length $\unitl$ is defined as $\unitl=\sqrt[3]{1/\rho}$ where $\rho$ is the number density of the particles.
Furthermore, we measure magnetic moments, velocities, and frequencies in multiples of $m_0=\sqrt{4\pi k \unitl^4/\mu_0}$, $k/c$, and $k/c \unitl$, respectively, with $c$ setting the viscous friction coefficient of each particle.\footnote{There is a typo in the definition of $m_0$ in Ref.~\cite{pessot2014structural}: it should read $m_0=\sqrt{4\pi k \unitl^5/\mu_0}$ instead of $m_0=\sqrt{4\pi k^2 \unitl^5/\mu_0}$.}
For our purposes, we assume $\sigma^{s}=0.2 \unitl$ and $\varepsilon^{s}=k \unitl$.

For reasons that will become clear in section \ref{static_elm_norm_modes}, it is useful to explicitly define and indicate the boundaries of our system.
We here consider samples of cubelike shape with faces perpendicular to $\dirx$,  $\diry$, and $\dirz$, the unit vectors defining our Cartesian coordinate system.
We can define ``left'' and ``right'', ``front'' and ``rear'', as well as ``bottom'' and ``top'' boundaries, namely the faces oriented by $\mp \dirx$, $\mp \diry$, and $\mp \dirz$, respectively.
The criteria to identify which particles belong to the boundaries will be detailed later according to the specific particle distribution.
Subsequently, we indicate by $L_x$, $L_y$, and $L_z$ the extension of the sample in the $x$-, $y$-, and $z$-direction, respectively.
In the case of cubelike shape and uniform density, $L_\alpha$ ($\alpha=x,y,x$) will be proportional to $N^{1/3}\unitl$.
Otherwise, an additional geometry-dependent prefactor can be included.
Then the scaling of cross-sectional areas (i.e.~$S_x=L_y L_z$) and the volume $V=L_x L_y L_z$ follow straightforwardly as $N^{2/3}\unitl^2$ and $N\unitl^3$, respectively.

\section{Equilibrium State}\label{equil_state}
First, we need to find the equilibrium state of our system, i.e.~the one that minimizes the total energy $U=U^{el}+U^{s}+U^{m}$ with respect to all degrees of freedom.
In our case the degrees of freedom are given by the positions $\vR_i$, which requires
\begin{equation}\label{equilR}
 \grad{U}{\vR_i} = \bm{0}, \ \ \ \ \forall \ i=1\dots N
\end{equation}
in equilibrium.
From Eqs.~(\ref{eel})--(\ref{emagn}) it is straightforward to calculate the resulting gradients (see Appendix \ref{der_app}).
The second derivatives of the energy $U$ form the corresponding Hessian matrix, see below.
Analytical expressions are listed in Appendices \ref{der_app} and \ref{hess_app}.

We seek the minimum total energy $U$ of a sample composed of $N$ particles arranged according to a prescribed distribution, each carrying a prescribed magnetic dipole moment $\vm$.
Consequently, the equilibrium state is obtained as a function of $\vm$.
To ease the convergence of the minimization techniques, we gradually increase the magnitude of the magnetic moments from $m=0$ (ground state) to the required maximum value of $m$ while minimizing the total energy for each intermediate value of $m$.
Because of the large number of degrees of freedom, the only practical way to find the equilibrium state is to perform a numerical minimization of the energy.
In the present work we implemented a conjugated gradient algorithm with guaranteed descent \cite{hager2005new}.

We wish to study the dynamic response of our systems for different orientations while holding $\vm$ fixed in space.
However, once the orientation of the magnetic moments is fixed from outside, the system as a whole may start to rigidly rotate to minimize its overall energy.
In real samples, such rotations are for instance suppressed by macroscopic frictional and gravitational forces.
Moreover, in our previous investigation, this macroscopic rotation was hindered by a ``clamping'' protocol of the boundaries \cite{pessot2014structural}.
Here instead, we develop a new protocol to keep the system in the desired orientation.
This is achieved by subtracting from the force field acting on the boundaries those parts corresponding to rigid rotations (see below and Appendix \ref{torq_app}).
This way, three constraints are applied in the form of the suppressed rigid rotations and we otherwise allow a complete internal relaxation of the sample.

\section{Normal Modes}\label{norm_modes}

Next, we describe a generic normal mode formalism and explain how it can be employed to characterize the linear response of our systems to a small external perturbation.
We do not assume regular, periodic particle distributions.
Instead, our formalism can likewise be applied to irregular particle arrangements, see, e.g., Refs.~\cite{shintani2008universal,rovigatti2011vibrational,lerner2014breakdown}.

In the following, we  indicate with a bra-ket notation $\vX$, the $D$-component vector containing all the $D$ degrees of freedom of the system.
In our case, $D=3N$ as we only consider translational degrees of freedom, but in principle $\vX$ could also include, for instance, particle rotations.

Once we write down the total energy $U(\vX)$, the equilibrium state $\vX^{eq}$ is given by the condition
\begin{equation}\label{equil}
 \grad{U}{\bm{X}}\left( \vX^{eq} \right) = \bm{0}.
\end{equation}
It is more convenient to discuss the problem in terms of displacement from equilibrium, $\vu = \vX -\vX^{eq}$.
Furthermore, it is always possible to shift the energy by a constant so that $U(\vX^{eq})=0$.
Around its minimum, we can expand $U(\vX)$ to lowest order in the displacement $\vu$:
\begin{equation}\label{taylor}
 U\left(\vu\right) \simeq \frac12 \vut  \hess  \vu,\ \mbox{with}\ \hess_{ij} = \partial_{{u}_i}\partial_{{u}_j} U.
\end{equation}
Here, $\hess$ is the Hessian matrix composed of the second derivatives of $U$ with respect to $\vu$ (see Appendices \ref{der_app} and \ref{hess_app}).
If $U(\vX)$ has continuous second partial derivatives, then $\hess$ is symmetric.
Moreover, being in a minimum of $U(\vX)$ implies that $\hess$ is positive-semidefinite.
All its eigenvalues are positive, except for the modes representing rigid translations and rotations, which cost no energy and have vanishing eigenvalues.

We obtain the linearized gradient around the minimum from Eq.~(\ref{taylor}) as
\begin{equation}\label{lingrad}
 \grad{U}{\bm{u}}\left(\vu\right) \simeq \hess \vu.
\end{equation}
When a small external force $\vf$ is applied, the system reacts to neutralize it and re-equilibrates:
\begin{equation}\label{response}
 -\grad{U}{\bm{u}}(\vu) + \vf = 0 \ \Rightarrow \ \hess \vu\simeq \vf.
\end{equation}
In Eq.~(\ref{response}) we have used Eq.~(\ref{lingrad}), which is justified for small $\vf$.
We diagonalize $\hess$ and introduce its eigenvalues $\lambda_n$ and eigenvectors, i.e.\ the normal modes $\vv{n}$ with $n=1\dots D$ and $D$ the number of degrees of freedom, such that
\begin{equation}\label{eigen}
 \hess \vv{n} = \lambda_n \vv{n}, \ \mbox{and} \ \scal{\bm{v}_m}{\bm{v}_n} = \delta_{mn},
\end{equation}
where $\delta_{mn}$ is the Kronecker delta.
Since the $\vv{n}$ form a complete basis, we can expand displacements and forces as
\begin{equation}\label{expans}
\vu = \sum_{n=1}^D u_n \vv{n} \ \mbox{and} \ \vf = \sum_{n=1}^D f_n \vv{n}.
\end{equation}
Here, $u_n=\scal{\bm{u}}{\bm{v}_n}$ and  $f_n=\scal{\bm{f}}{\bm{v}_n}$.
Then, using these expansions and the orthonormality of the eigenvectors, Eq.~(\ref{response}) simply reduces to
\begin{equation}\label{lresponse}
 \lambda_n u_n = f_n.
\end{equation}
This relation clearly shows that, under the influence of an external force $\vf$ exciting the $n$-th normal mode, the amplitude $u_n$ of the response is linearly related to the intensity $f_n$ of the force.
In this perspective, the Hessian eigenvalue $\lambda_n$ quantifies the magnitude of the static linear response of the system within the $n$th mode to the external force.
$\lambda_n$ is therefore a sort of elastic constant.
Thus, the energy of the system around its minimum can be written, using Eqs.~(\ref{taylor}), (\ref{expans}), and (\ref{lresponse}), as
\begin{equation}\label{Umodes}
 U = \frac12 \sum_{n=1}^D \lambda_n {u_n}^2= \frac12 \sum_{n=1}^D \frac{{f_n}^2}{\lambda_n}.
\end{equation}

\section{Static Elastic Moduli from Normal Modes}\label{static_elm_norm_modes}
In numerical calculations there are two main ways to obtain elastic moduli in the zero-frequency limit, i.e., in the static case.
On the one hand, one can perform a finite but small (linear-regime) strain of the whole system, both for \pbc\ \cite{frenkel2002understanding,frenkel1987elastic,weeber2015ferrogels2} or \obc\ \cite{pessot2014structural}.
The system is equilibrated under the prescribed amount of strain.
In this way, the moduli are measured from the slope of the resulting stress-strain curve or, equivalently, from the second derivatives of the free energy.
On the other hand, when employing \pbc\ and working in thermodynamic equilibrium, one can differentiate the free energy with respect to a macroscopic strain \cite{squire1969isothermal,frenkel2002understanding, born1939thermodynamics}.
As a special case, and in the low-temperature limit, the elastic moduli of a \pbc\ glassy system have recently been examined \cite{fuereder2015influence}, whereas the case of regular lattices was discussed under the assumption of affinity in the deformation \cite{ivaneyko2012effects}.
However, it is important to remark that affinely mapping the macroscopic strain down to all scales in the system does not allow for internal relaxation \cite{jaric1988density} and can even lead to qualitatively incorrect results \cite{pessot2014structural} in presence of non-affinity sources.

In the present work we consider the case of a finite system in the ground state neglecting thermal fluctuations of the mesoscopic particles.
The semi-analytical approach that we use to calculate elastic moduli in the linear regime does not require finite macroscopic displacements nor does it assume affinity of the deformation.
This method relies on the decomposition of the linear response over the eigenvectors of the Hessian matrix $\hess$.
It reduces the calculation to a problem of linear algebra and gives access to dynamic properties as well, see sections \ref{dynamics} and \ref{dynamic_elm}.
Physically, our procedure involves using stress instead of strain as an independent variable.

\subsection{Macroscopic Stresses and Strains}\label{macr_stress_strain}
Below we will focus on Young's modulus $E$ and the shear modulus $G$.
They can be defined via the stress-strain relationships:
\begin{equation}\label{stress-strain}
\sigma_{\alpha\alpha} = E_{\alpha\alpha}\ \varepsilon_{\alpha\alpha}, \ \mbox{   } \ \sigma_{\alpha\beta} = G_{\alpha\beta}\ \varepsilon_{\alpha\beta},
\end{equation}
\GP{where $\sigma_{\alpha\beta}$ ($\alpha,\beta=x,y,z$) denotes the force per area applied in the $\beta$-direction acting on the boundary with the surface normal oriented in the $\alpha$-direction.
$\varepsilon_{\alpha\beta}$ indicates the corresponding strain deformation, i.e.\ the total displacement of the boundary in the $\beta$ direction divided by the distance between the boundaries in the $\alpha$-direction.}
Here, there is no summation over $\alpha$ \GP{and $\beta$}.
In the first formula, $\alpha$ defines the direction of imposed stretching or compression, along which we evaluate $E_{\alpha\alpha}$.
In the second formula, the $\alpha\beta$ plane sets the shear plane within which we evaluate $G$, with the shear displacement on the boundaries introduced along the $\beta$-direction.
Thus, only the faces of the system perpendicular to the $\alpha$-direction need to be explicitly addressed to impose our boundary stresses, while the rest of the system is free to relax.
This configuration conceptually reproduces an experimental situation in which the sample would be enclosed between the plates of a rheometer with the plates perpendicular to the $\alpha$-direction \cite{collin2003frozen}.

\GP{Applying during shear only forces oriented tangential to the surface planes typically induces rotations.
In experiments, these are hindered by the confining plates.
Accordingly, we here suppress such global rotations by subtracting them from the overall response of the system (see below and appendix \ref{torq_app}).
In this way, we maintain the definition of $\sigma_{\alpha\beta}$ as above close to the experimental situation and avoid symmetrization typically performed in the context of classical elasticity theory \cite{landau1975elasticity} (for a related discussion on anisotropic systems see also Ref.~\cite{menzel2014bridging}).}

In the following derivation, we focus on the Young modulus $E_{\alpha\alpha}$ and drop the $_{\alpha\alpha}$ subscripts.
The calculation for the shear modulus $G_{\alpha\beta}$ is analogous.
Here, stresses and strains in Eq.~(\ref{stress-strain}) are interpreted as macroscopic quantities characterizing the overall deformation of the system.
We measure them and accordingly define the elastic moduli of the system solely by the stresses on and the displacements of the boundaries perpendicular to $\hat{\bm{\alpha}}$, respectively.
The stress is calculated from the ratio between the external force and the surface over which it is applied.
Similarly, the strain is obtained by measuring the displacement of the boundaries and dividing by their initial distance.

The energy of a strain deformation is given by the work performed by the stress in the whole volume, i.e., using Eq.~(\ref{stress-strain}),
\begin{equation}\label{str_en}
 U = V \int \sigma\ \d \varepsilon\ = V \frac{E\varepsilon^2}{2} = V \frac{\sigma^2}{2E}.
\end{equation}
Therefore, the elastic modulus can be derived by differentiating the previous equation,
\begin{equation}\label{macr_elm}
 E = \frac{1}{V} \stder{U}{\varepsilon}= V {\left[ \stder{U}{\sigma} \right] }^{-1}.
\end{equation}

\subsection{Mesoscopic Stress}\label{micr_stress}
Our goal is to connect these macroscopic relations to the mesoscopic level.
On the mesoscopic scale, within our linear response framework, it is impractical to use the strain as a variable to impose an external perturbation of the system.
Imposing a certain amount of strain by displacing the boundary particles in a prescribed way does not provide any information on the displacement of the bulk particles because the internal relaxation of the system is not known \textit{a priori}.
Actually, the rearrangement of the bulk particles mainly determines the reaction of the system and contributes the most to the elastic response.
In contrast to that, it is more convenient to use the stress as a variable to impose the external perturbation when we connect the macroscopic to the mesoscopic level.
As a matter of fact, we know that an externally imposed mechanical stress leads to nonvanishing external forces on the boundary particles only.

\begin{figure}[]
\centering
  \includegraphics[width=8.6cm]{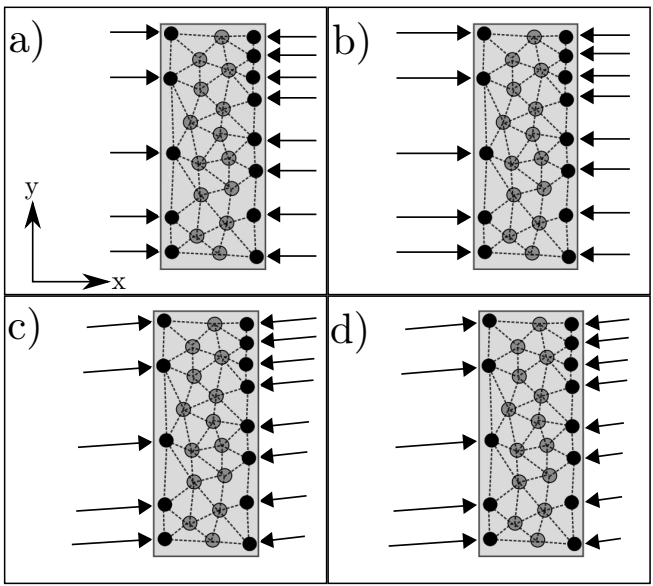}
  \caption{Protocol to connect a macroscopic stress ($\sigma_{xx}$) acting on the system boundaries to a discretized mesoscopic force field acting on the boundary particles.
For simplicity, the case of an irregular two-dimensional ($2$D) system is shown here.
Particles on the boundaries are colored in black and springs are represented by dotted lines.
This figure is for illustrative purposes only, therefore lengths and vectors are scaled in a qualitative way.
Our procedure is as follows:
a) First, individual discrete forces of equal magnitude are introduced on each individual boundary particle, pointing into the stress-direction (here the $x$-direction).
b) The forces are rescaled to balance total forces on the left- and right-hand sides.
c) An appropriate rotatory component is introduced to make the torques vanish on each boundary (separately).
d) All forces on each boundary are rescaled by a common factor so that their sum in the stress-direction is normalized correctly.}
  \label{fig_micr_stress}
\end{figure}

We here describe the macroscopic mechanical stress $\sigma$ in terms of sets of discretized forces acting directly on the mesoscopic particles.
We denote the number of particles on the ``left'' and ``right'' boundaries (see section \ref{dip_spring_model}) as $N_l$ and $N_r$, respectively.
If we indicate by $S$ the cross-section over which a total external force $F$ is applied, then we have $F=\sigma S$.
The corresponding externally imposed discretized mesoscopic force field $\vf$ acting directly on the particles can then be constructed using the following protocol:
\begin{enumerate}[label={\alph*)}]\label{force_cond}
 \item $\vf$ is non-vanishing only on the boundaries and has components oriented in the stress-direction, see Fig.~\ref{fig_micr_stress} a).
 \item The total force $F$ acting on one boundary must be equal in magnitude to the total force acting on the other boundary.
First, we assume all individual forces acting on individual particles on the same boundary to be equal in magnitude.
We indicate by $f_l$ and $f_r$ those forces acting on a single individual particle on the left or right boundary, respectively.
Then the condition reads $F=N_l f_l = N_r f_r$, see Fig.~\ref{fig_micr_stress} b).
 \item The torque exerted by $\vf$ on the boundaries must vanish [see Fig.~\ref{fig_micr_stress} c)].
This can be achieved using the method described in Appendix \ref{torq_app}.
The condition is applied separately to each boundary.
\item Finally, we must rescale all forces acting onto one boundary by a common factor so that the forces acting in the stress direction sum up to $F=\sigma S$ [see Fig.~\ref{fig_micr_stress} d)].
Again, this condition is applied separately to each boundary.
\end{enumerate}
These steps serve as a protocol when generating the discretized boundary force field $\vf$ in numerical calculations.
In the following, we factor out $F$ and write $\vf = \sigma S \vfnrm$, where $\vfnrm$ is a force field satisfying our requirements and representing a macroscopic force of unitary magnitude ($F=1$).

\subsection{Calculation of Static Elastic Moduli}\label{stat_elm_calc}
We now have all ingredients available to formulate the connection between the macroscopic elastic modulus and our discretized mesoscopic normal modes.
Following the definition of particle-resolved stress $\sigma S \vfnrm$ that we introduced above, we write the energy in Eq.~(\ref{Umodes}) as an explicit function of $\sigma$,
\begin{equation}\label{energy_stress}
 U = \frac{\sigma^2 S^2}{2} \sum_{n=1}^D \frac{ {{{f}_n^u}}^2 }{\lambda_n}, \ \mbox{with} \ {f}_n^u = \scal{{\bm{f}^u}}{\bm{v}_n}.
\end{equation}
Combining it with Eq.~(\ref{macr_elm}), we obtain
\begin{equation}\label{stat_elmod}
 E = \frac{L}{S} {\left[  \sum_{n=1}^D \frac{ {{{f}_n^u}}^2 }{\lambda_n} \right]}^{-1}.
\end{equation}
Here, again, $S$ is the surface area of the boundary on which the stress acts, while $L$ is the distance between the two boundaries so that $LS=V$.
$\lambda_n$ is the $n$-th eigenvalue of the Hessian matrix, and ${f}_n^u$ is given by Eq.~(\ref{energy_stress}).
In general, $S$ and $L$ will be proportional to $N^{(d-1)/d}\unitl^{d-1}$ and $N^{1/d}\unitl$, respectively, with $d$ the \GP{spatial} dimensionality of the system.
Therefore, for $3$D particle arrangements of cubelike shape we obtain $L/S \sim 1/\sqrt[3]{N} \unitl$.
In other cases a prefactor must be added, taking into account the shape of the sample or the unit cell structure in the case of regular lattices.

In the following numerical calculations we used the \textsc{lapack} diagonalization routines \cite{anderson1999lapack} to find eigenvalues and eigenvectors of $\hess$.
Special care must be taken to avoid the zero-energy modes when computing Eq.~(\ref{stat_elmod}). We here simply ignore contributions from the lowest $3$ and $6$ eigenvalues when dealing with $2$D and $3$D systems, respectively.
They correspond to rigid translations and rotations of the system.

Overall, we have described a self-standing procedure to calculate elastic moduli in \obc\ systems.
The system is required to be in a stable equilibrium state, where the Hessian matrix of the total energy is positive semi-definite.
Since the elastic moduli are properties of the ground state, they can be directly obtained via the eigenvalues and eigenvectors calculated in this configuration, see Eq.~(\ref{stat_elmod}), for a specified force field, see section \ref{micr_stress}.
Therefore, it is not necessary to actually perform a finite deformation and drive the sample out of equilibrium as e.g.\ in Refs.~\cite{frenkel1987elastic,weeber2015ferrogels2,pessot2014structural}.
In the following section we compare the results of our described method with those obtained by explicitly taking a system out of equilibrium via actual boundary displacement.

\subsection{Comparison with $2$D Calculations}
The calculation we outlined in section \ref{stat_elm_calc} has the advantage of requiring knowledge of only the ground state to obtain all (linear) elastic moduli.
Conversely, as we just mentioned, the previously taken path to determine the elastic moduli is to drive the system out of the ground state by prescribing a small amount of strain, determining its deformation, and thereby tracking the total energy variations, see e.g.\ Ref.~\cite{pessot2014structural,weeber2015ferrogels2,cremer2015tailoring}.
To test the validity of the present approach, we compare the method described above with the numerical results obtained \GP{previously} for the $2$D case via explicit boundary displacements \cite{pessot2014structural}.

\begin{figure}[h]
\centering
  \includegraphics[width=8.6cm]{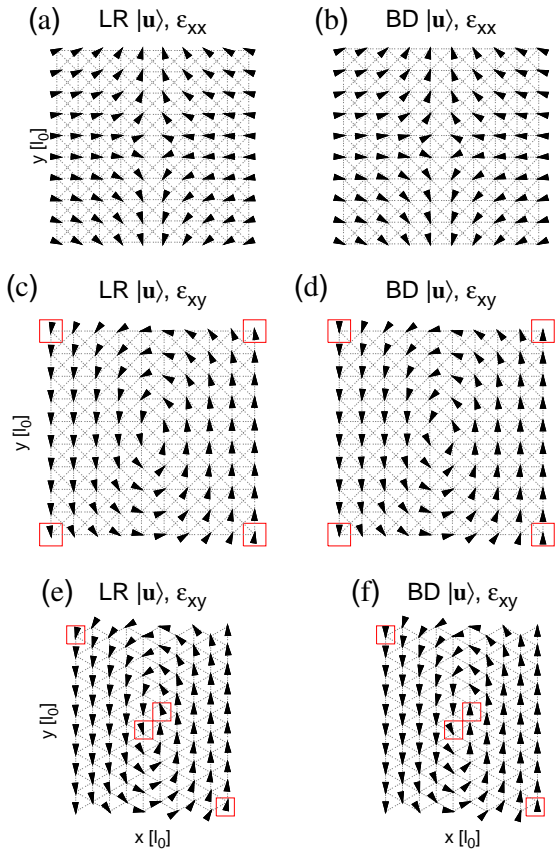}
  \caption{Non-affine displacement field $\vu$ of exemplary square and triangular lattices composed of $100$ particles (springs indicated by dashed lines) for $m=0$ obtained with \lr\ and \bd\ methods [panels (a), (c), (e) and (b), (d), (f), respectively] for \GP{stretching/compression} $\varepsilon_{xx}$ and simple shear $\varepsilon_{xy}$ deformations [panels (a), (b) and (c), (d), (e), (f), respectively].
This simple, exemplary case shows how the responses obtained from the two methods are both non-affine and similar, but can yet present small differences (compare e.g.\ particles highlighted by red squares), explaining small deviations in the elastic moduli resulting from the two methods, see Fig.~\ref{fig_e_m_comp}.
Panels (b), (d), (f) were obtained by imposing small (linear-elasticity regime) strains of $\varepsilon_{xx}=0.03$ and $\varepsilon_{xy}=0.001$, respectively.}
  \label{fig_lin_resp_comp}
\end{figure}

We briefly sum up the technique applied in our former work, see Ref.~\cite{pessot2014structural}.
In that case, a $2$D dipole-spring model, similar to the present one but without steric repulsion, is considered.
The left and right boundaries of the system are set perpendicular to the $x$-direction and undergo a ``clamping'' protocol, i.e., all the particles in the boundary are constrained to move along $\dirx$ or $\diry$ in a prescribed way and therefore the whole system undergoes a determined amount of strain $\varepsilon_{xx}$ or $\varepsilon_{xy}$.
For every prescribed position of the boundaries, the bulk of the system is free to relax [see Fig.~\ref{fig_lin_resp_comp} (b), (d), and (f)].
Then, the static Young's modulus is obtained from the second derivative of the total energy with respect to a small strain in the linear elasticity regime.

Contrarily to the present case, in Ref.~\cite{pessot2014structural} we considered springs of identical elastic constant, regardless of the spring length.
To allow a better comparison with our former results we will---solely in this subsection---assign an equal elastic constant to all springs, i.e.\ $k_{ij} = k$ $\forall i,j$.
Moreover, for the present $2$D setup, the elastic moduli will be measured in multiples of $k$.
In the following, we will address the previous calculations of Ref.~\cite{pessot2014structural} as ``Boundary Displacement'' (\bd) and those in the framework of linear response theory of the present work \GP{as} ``Linear Response'' (\lr).

{We first consider the case of a $2$D square spring lattice with nonmagnetized ($m=0$) particles on the vertices.
On the one hand, and in the \bd\ case, we can apply a prescribed, small amount of strain $\varepsilon_{xx}$ or $\varepsilon_{xy}$ and, after full internal energetic relaxation, observe the resulting displacement field \bd\ $\vu$, see Fig.~\ref{fig_lin_resp_comp} \GP{(b), (d), and (f)}.
On the other hand, and in the present \lr\ scheme, we start from the small mesoscopic force field $\vf$ as constructed via the protocol described in section \ref{micr_stress}.
The corresponding coefficients $f_n$ are obtained from Eq.~(\ref{expans}).
Then, using the eigenvalues of the Hessian matrix $\lambda_n$ as well as Eq.~(\ref{lresponse}), we obtain the response of the modes, i.e.\ the coefficients $u_n$.
Finally, using the coefficients $u_n$, we obtain  via Eq.~(\ref{expans}) the particle-resolved displacement \lr\ $\vu$, which is the linear response of the system to the small applied force $\vf$, see Fig.~\ref{fig_lin_resp_comp} \GP{(a), (c), and (e)}.

The comparison between the resulting displacement fields is helpful to understand where small deviations between the elastic moduli obtained via the two different methods may arise from, see Fig.~\ref{fig_e_m_comp}.
Overall, the differences remain small, especially in the case of \GP{stretching and compression} [see Fig.~\ref{fig_lin_resp_comp} (a) and (b)].
For shear deformations [see Fig.~\ref{fig_lin_resp_comp} (c) and (d)], such discrepancies are visible and reflect small deviations in the resulting moduli.
This effect seems to be stressed when the positions of boundary particles are not mirror symmetric with respect to the direction of the calculated modulus, as in the case of the triangular lattice for Young's modulus in $x$-direction in Fig.~\ref{fig_lin_resp_comp} (e) and (f).
In total, however, we may conclude that our protocol to construct the force field, see section \ref{micr_stress}, works well and reproduces the mesoscopic displacement fields previously obtained via \bd.

To further test the performance of the present method, we now consider magnetic particles ($m\neq 0$).
We compare some of the elastic moduli obtained in Figs.~5, 6, and 7 of Ref.~\cite{pessot2014structural} as functions of $m$ for a few exemplary cases of regular lattice structures.
As shown in Fig.~\ref{fig_e_m_comp}, we find the same behavior for $E(m)$ depending on lattice structure and neighbor orientation.
\begin{figure}[]
\centering
  \includegraphics[width=8.6cm]{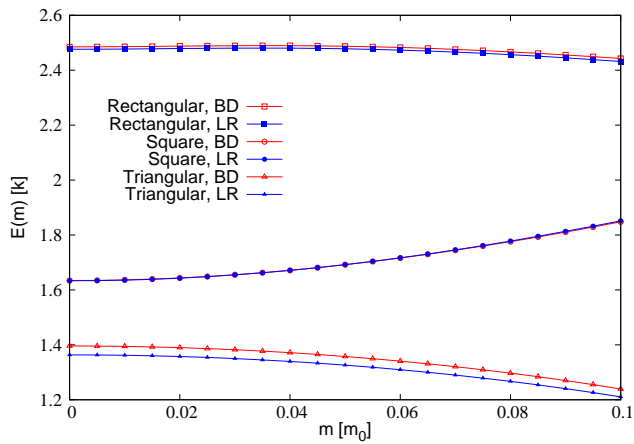}
  \caption{Young's modulus $E$ as a function of $m$ calculated for comparison with \bd\ and \lr\ techniques.
Three cases are presented (top to bottom): rectangular lattice of base-height ratio $b/h=2.5$, square lattice ($b/h=1$), and triangular lattice [see panels (e) and (f) of Fig.~\ref{fig_lin_resp_comp}] with $\vm$ oriented in the $x$-, $z$-, and $y$-direction, respectively.
The number of particles in all of the three examples is $N=400$.
The triangular lattice case shows a comparatively larger difference, which, however, does not depend on $m$.
We mostly attribute such deviations to the structure of the boundary, as detailed in Fig.~\ref{fig_lin_resp_comp} (e) and (f).}
  \label{fig_e_m_comp}
\end{figure}
Depending on the particle arrangement, small discrepancies can appear, as explained above.
These deviations also seem to depend on the specific shape of the boundaries and are more evident for the case of the triangular lattice in Fig.~\ref{fig_lin_resp_comp} (e) and (f).
From now on, we will turn back to the more general $3$D case.

\section{Dynamics}\label{dynamics}
Because of their \GP{often highly} viscous character \GP{on the mesoscale}, soft matter systems in motion \GP{typically} undergo large dissipation and their dynamics is studied in the overdamped regime \cite{ivlev2012complex,tarama2014tunable,ivaneyko2015dynamic,weeber2015ferrogels2}.
In the following we describe the time-evolution of our systems, starting from the overdamped equation of motion.
Then, a way to decouple the full equation of motion in the normal modes is presented and the general solution for a single mode is shown.

To keep the derivation general, we here take up the notation introduced in section \ref{norm_modes} with the difference that now $\vu(t)$ and $\vf(t)$ depend on time.
The full, coupled equation of motion for the overdamped dynamics of the system can be written as
\begin{equation}\label{eq_motion}
 \mathcal{C} \left|\dot{\bm{u}}\right\rangle(t) + \mathcal{H} \vu (t) = \vf (t),
\end{equation}
where the dot represents time differentiation, the matrix $\mathcal{C}$ contains the (viscous) friction coefficients, and we have used the linearized version of the gradient $\mathcal{H} \vu$ as in Eq.~(\ref{lingrad}).
Here, for simplicity and as a first step, we consider the case of mesoscopically isotropic building blocks under negligible long-ranged dynamic coupling, i.e.\ $\mathcal{C}= c \mathbb{I}$, with $\mathbb{I}$ the $D\times D$ identity matrix and $c$ the viscous friction coefficient for one isotropic particle.

As a consequence, the matrices $\mathcal{C}$ and $\mathcal{H}$ commute and can be simultaneously diagonalized, i.e.~they have a common base of eigenvalues, namely the $\vv{n}$ in Eq.~(\ref{eigen}).
Then, using the normal modes, Eq.~(\ref{eq_motion}) of $D$ variables can be decoupled into $D$ independent single-variable equations
\begin{equation}\label{damped_ho}
 c \dot{u}_n (t) + \lambda_n u_n (t) = f_n (t),
\end{equation}
with $n=1\dots D$.
If the external force $\vf(t)$ is periodic, i.e.~$\vf(t)=\left|\bm{f}^0\right\rangle\exp{\left(\imu\omega t\right)}$, its projections onto the Hessian eigenvectors $\vv{n}$ will be equally periodic,
\begin{equation}\label{force_per_decomp}
f_n(t)=f_n^0\exp{\left(\imu\omega t\right)},
\end{equation}
with $f_n^0 = \scal{\bm{f}^0}{\bm{v}_n}$.
Thus, the solution $u_n (t)$ of Eq.~(\ref{damped_ho}) \GP{after all transients have decayed} must be periodic as well, i.e.\
\begin{equation}\label{osc_displ}
u_n(t)=u_n^0\exp{\left(\imu\omega t\right)}.
\end{equation}
Substituting the last equations into Eq.~(\ref{damped_ho}), we obtain
\begin{equation}
u_n^0 = f_n^0/\kappa_n(\omega)
\end{equation}
with
\begin{align}\label{kappa}
\kappa_n(\omega) &= \lambda_n +\imu c \omega\\
&= \e^{\imu\delta_n(\omega)}\lambda_n\sqrt{1+{\tau_n}^2\omega^2}, \nonumber \\
\mbox{where} \ \delta_n(\omega) &= \arctan{\left( \tau_n\omega\right)}. \nonumber
\end{align}
In these expressions we introduced by $\tau_n=c/\lambda_n$ the relaxation time and by $1/\kappa_n(\omega)$ the dynamic linear response function of the $n$-th mode.

As described above, we focus on the overdamped dynamics and do not include inertial terms in Eq.~(\ref{eq_motion}).
If an inertial term had been considered, it would have resulted in a ${(\lambda_n -\widetilde{m} \omega^2)}^2$ term inside the square root of Eq.~(\ref{kappa}), with $\widetilde{m}$ the mass of one particle.
Such a contribution would have showed up as a resonance frequency $\tilde{\omega}_n = \sqrt{\lambda_n/\widetilde{m}}$ for the $n$-th mode.
As a consequence, when the frequency of the driving force $\omega$ coincides with $\tilde{\omega}_n$, large displacements can be induced by small external perturbations.
Such an effect would result in a significant drop of the elastic moduli at frequencies close to the resonances of those modes that contribute most to the linear response.
This behavior, however, is not obvious from experimental reports \cite{zanna2002influence,sedlacik2016magnetorheological,roeben2014magnetic,hohlbein2015remote}, thus supporting the overdamped approach.
Eq.~(\ref{kappa}) implies that the displacement $u_n(t)$, i.e.\ the response, chases the driving force $f_n(t)$ with identical frequency.
However, because of viscous friction, it follows with a phase lag $\delta_n(\omega)$, which vanishes in the case of frictionless motion.
Such a phase lag implies an imaginary component of $\kappa_n(\omega)$ corresponding to a loss component of the elastic moduli, see below.

\section{Dynamic Elastic Moduli}\label{dynamic_elm}
We aim at extending the normal modes treatment that we carried out for Eq.~(\ref{response}) and transfer it to the dynamic situation described by Eq.~(\ref{eq_motion}).
The final goal will be to generalize Eq.~(\ref{stat_elmod}) for the macroscopic overall elastic moduli to the case of periodically oscillating external stresses and thus obtain the dynamic elastic macroscopic moduli.
We here consider the case of a Young modulus $E(\omega)=E_{\alpha\alpha}(\omega)$ for direction $\alpha \in \{x,y,z\}$.
The discussion of a shear modulus $G_{\alpha\beta}(\omega)$ is entirely analogous, provided that the protocol prescribed in section \ref{micr_stress} is followed.

We now start with a macroscopic, periodic, and single-frequency stress
\begin{equation}
\sigma(t) = \sigma^0\e^{\imu\omega t}
\end{equation}
applied to the sample, with $\sigma^0$ a real amplitude.
The resulting macroscopic strain $\varepsilon(t)$ varies with the same frequency.
Thus we write
\begin{equation}
\varepsilon(t) = \varepsilon^0(\omega)\e^{\imu\omega t},
\end{equation}
where $\varepsilon^0(\omega)$ is, in general, a complex amplitude.
Using these expressions in the single-frequency case, the frequency-dependent dynamic modulus $E(\omega)$ follows via
\begin{equation}\label{dyn_stress_strain}
\sigma(t) = E(\omega) \varepsilon(t) \ \ \Leftrightarrow \ \ E(\omega) = \frac{\sigma^0}{\varepsilon^0(\omega)}.
\end{equation}
Thus, $E(\omega)=E'(\omega) +\imu E''(\omega)$ has an imaginary part whenever $\sigma(t)$ and $\varepsilon(t)$ are not completely in phase and can be divided into storage ($E'$) and loss ($E''$)
components.

Now we take up again the formalism of sections \ref{norm_modes} and \ref{static_elm_norm_modes}.
On the mesoscopic level---see section \ref{dynamics}---the time-dependent response $\vuo \exp{\left(\imu\omega t\right)}$ of the system, \GP{after all transients have decayed,} is related to a {small} driving force $\vfo \exp{\left(\imu\omega t\right)}$ by
\begin{equation}\label{dyn_response}
\vuo \e^{\imu\omega t} = \sum_{n=1}^{D} u_n^0 \vv{n}\e^{\imu\omega t} = \sum_{n=1}^{D} \frac{f_n^0}{\kappa_n(\omega)} \vv{n}\e^{\imu\omega t},
\end{equation}
where, again, $D$ is the number of degrees of freedom, $f_n^0=\scal{\bm{f}^0}{\bm{v}_n}$, $u_n^0=\scal{\bm{u}^0}{\bm{v}_n}$, and we used Eq.~(\ref{expans}).

The macroscopic dynamic stress is given by $\sigma(t)=F\exp(\imu \omega t)/S$, with $S$ the boundary surface area and $F$ the macroscopic force acting on it.
Moreover, the macroscopic strain is $\Delta/L$ with $\Delta$ the change in separation of the macroscopic sample boundaries and $L$ the absolute distance between them.
The displacement $\Delta$ is measured in the direction of the applied force inducing it.
Therefore, and since $\vfnrm$ represents the mesoscopic direction of a force of magnitude unity ($F=1$, see section \ref{micr_stress}), we define $\Delta=\scal{\bm{f}^u}{\bm{u}}$ as a measure of the resulting displacement.
We recall here that $\vfo$ was constructed to apply only \GP{on} the boundary, so $\scal{\bm{f}^u}{\bm{u}}$ really extracts the displacement of the boundaries.
Consequently, we write Eq.~(\ref{dyn_stress_strain}) on the mesoscopic level as
\begin{equation}\label{dym_meso_stress_strain}
\frac{F\e^{\imu\omega t}}{S} = E(\omega) \frac{\scal{\bm{f}^u}{\bm{u}^0}\e^{\imu\omega t}}{L}.
\end{equation}
Using Eq.~(\ref{dyn_response}), as well as $f_n^0=F f_n^u$ and $f^u_n=\scal{\bm{f}^u}{\bm{v}_n}$ (see section \ref{micr_stress}), the dynamic modulus follows as
\begin{equation}\label{dyn_elm}
 E(\omega) = \frac{L}{S} {\left[\sum_{n=1}^D \frac{ {f_n^u}^2 }{\kappa_n(\omega)} \right]}^{-1}
\end{equation}
which does not depend on the macroscopic force intensity $F$ and in the case $\omega=0$ recovers Eq.~(\ref{stat_elmod}).
Since $\kappa_n(\omega)$ is a complex number, $E(\omega)$ is complex as well and we can separate it into storage and loss components $E(\omega)=E'(\omega) +\imu E''(\omega)$.
We remark that in the static case we always find $E''(\omega=0)=0$ by definition [see Eq.~(\ref{kappa})].

On the macroscopic level, Eq.~(\ref{dyn_elm}) is connected to the Kelvin-Voigt model, which correctly describes the properties of permanently crosslinked polymers on long times scales, i.e., small $\omega$.
This is clear in a limit case when a single mode, e.g.\ $n=1$, has a relaxation time, \GP{e.g.}\ $\tau_1=c/\lambda_1$, much longer than the other modes.
Then, the \GP{long-frequency dynamics} is dominated by this mode which gives, in fact, the largest contribution to the sum in Eq.~(\ref{dyn_elm}).
Eventually, in this case one would find $E(\omega) \propto \kappa_1(\omega) = \lambda_1 + \imu \omega c$, which is precisely the form of the dynamic modulus in the Kelvin-Voigt model \cite{mainardi2011creep,eldred1995kelvin}.

In the following, we will apply the present approach to different particle distributions, addressing the dynamic elastic moduli for varying $\omega$ and $m$.
Although we will display the behavior of the dynamic moduli up to relatively large values of $\omega$, one should keep in mind our focus on overdamped motion.
At maximum our approach is meaningful up to a frequency $\omega_{max}=\lambda_{max}/c$, where $\lambda_{max}$ is the largest eigenvalue of $\hess$.

The limit becomes visible from calculating the spectrum, i.e.\ \GP{the density of states $g(\omega)$ \cite{ashcroft1976solid}}.
It is defined by
\begin{equation}\label{dos}
{g}(\omega)= \frac1D \sum_{n=1}^D \delta\left( \omega -\frac{\lambda_n}{c} \right),
\end{equation}
with $\delta$ the Dirac delta function.
To determine it from our numerical calculations, we replace the Dirac delta function by a narrow normalized Gaussian.
We chose the Gaussians as narrow as possible to achieve a smooth representation of the density of states.
\begin{figure}[]
\centering
  \includegraphics[width=8.6cm]{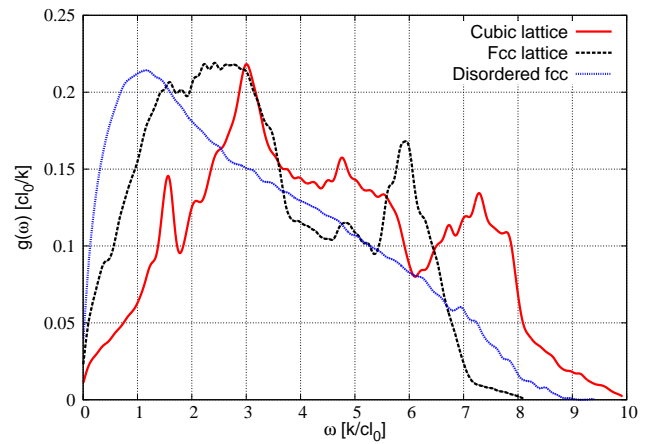}
  \caption{Density of states $g(\omega)$ at vanishing $m$ of a cubic lattice with springs up to second-nearest neighbors (see section \ref{cubic_result}), an \fcc\ lattice with only nearest-neighbor springs taken into account (see section \ref{fcc_result}), and a disordered lattice (see section \ref{disord_result}) made of $4913$, $6084$, and $6084$ particles, respectively.
The density of states is shown from $\omega=0$ to the highest $\omega_{max}$ obtained from the Hamiltonian spectrum, which is usually $\lesssim 10 k/c\unitl$.
The standard deviation of the narrow Gaussians used to approximate the Dirac deltas appearing in Eq.~(\ref{dos}) is chosen as $0.005\omega_{max}$.}
  \label{fig_dos}
\end{figure}

We always find $g(\omega)$ to drop significantly beyond a maximum value $\omega_{max}$.
The latter is of the order of a few $k/c\unitl$, see Fig.~\ref{fig_dos}.
Consequently, and because of our focus on the overdamped regime, it is not sensible to take into account the behavior for $\omega \gtrsim 10 k/c\unitl$.

First, the exemplary case of a simple cubic lattice will be studied.
After that, we consider an \fcc\ particle arrangement, before we finally move on to the case of disordered and more realistic particle arrangements.
For simplicity, we will always keep the magnetic moment $\vm$ oriented in the $z$-direction.
We measure the Young moduli in the perpendicular ($E_{xx}$ and $E_{yy}$) and parallel ($E_{zz}$) directions.
Likewise, the shear moduli will be calculated in the three possible orientations depicted in Fig.~\ref{fig_shear_sketch}: (a) shear corresponding to $G_{xy}$ does not \GP{directly} modify distances along the $\vm$-direction; (b) while $G_{xz}$ is measured the \GP{macroscopic} shear displacements are oriented along $\vm$; and (c) the shear plane contains $\vm$, but the \GP{macroscopic} shear displacements are perpendicular to $\vm$ when $G_{zy}$ is determined.
\begin{figure}[]
\centering
  \includegraphics[width=8.6cm]{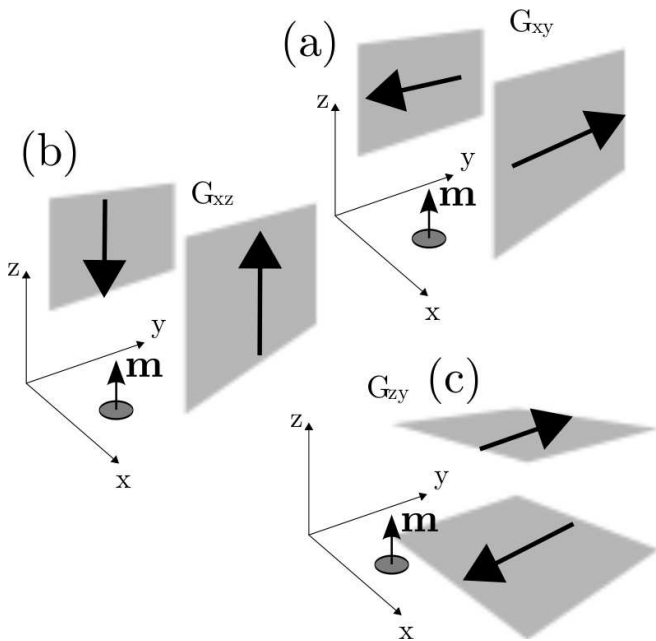}
  \caption{Illustration of the three principal shear geometries.
$\vm$ is rigidly oriented in the $z$-direction.
\GP{Shear forces can be applied to different boundaries and in different directions, giving origin to three main geometries (top to bottom):
(a) for $G_{xy}$ forces are perpendicular to $\vm$, but the \GP{driven boundary} planes contain $\vm$;
(b) for $G_{xz}$ \GP{both} shear forces \GP{and driven boundary planes} are parallel to the $\vm$ direction;
(c) for $G_{zy}$ the \GP{driven boundary} planes and shear forces are perpendicular to $\vm$.
We here define stresses directly via the forces acting on the indicated planes along the desired directions.}
}
  \label{fig_shear_sketch}
\end{figure}
Moreover, we here have $G_{yx}=G_{xy}$, $G_{xz}=G_{yz}$, and $G_{zx}=G_{zy}$.

\section{Cubic Lattice}\label{cubic_result}
As a first prototype, we consider the simple exemplary case of a $3$D cubic lattice with $N=\npcube$ particles.
Magnetic particles on the lattice are linked by springs up to second-nearest neighbors.
Corresponding springs along the diagonals of the faces of the unit cells are necessary to avoid unphysical soft-shear modes.
The boundaries of the system are simply identified as the outermost layers of particles in the respective directions.
As explained in section \ref{dip_spring_model}, the lattice parameter and the typical interparticle distance $\unitl$ follow from the number density $\rho$.
In the case of a simple cubic lattice structure, $\rho$ is given by one particle per unit cell.

Upon introducing a dipole magnetic moment in the particles, the direct attraction between nearest neighbors causes the system to shrink in the $\vm$-direction and expand in the perpendicular directions (see Fig.~\ref{fig_cubic_latt}).
\begin{figure}[]
\centering
  \includegraphics[width=8.6cm]{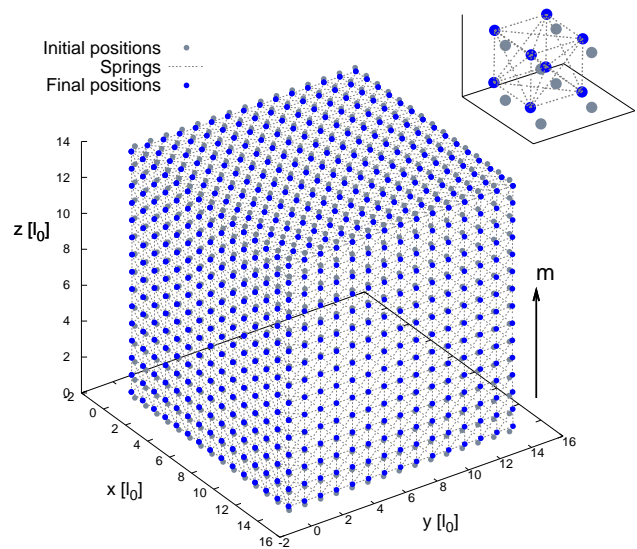}
  \caption{Deformation of an initially cubic lattice with springs between up to second-nearest neighbors and $N=\npcube$ when a magnetic moment of $\vm=0.1 m_0\dirz$ is gradually switched on.
For illustrative purposes, only particles on the front, top, and right faces are depicted.
Shrinking is observed along $\vm$, i.e.\ the $z$-direction, and dilation in the perpendicular directions.
The inset zooms in onto the deformation of the unit cell at the bottom left corner of the sample.}
  \label{fig_cubic_latt}
\end{figure}
Technically, in our numerical calculations, we gradually increased the magnetic moment to the value under consideration, up to a maximum magnitude of $m=0.1m_0$.
In this regime, and despite the overall deformation, the lattice maintains a cuboidlike shape.
The magnetic interactions are not as strong as to overcome the elastic springs and the particles do not come into steric contact.

\subsection{Static moduli}\label{sc_result_static}
We start by studying the static moduli $E$ and $G$ (i.e.\ the storage components $E'$ and $G'$ of the dynamic moduli calculated for $\omega=0$) for increasing magnitude of the magnetic moment $m$, see also Ref.~\cite{ivaneyko2012effects}.
Magnetic interactions between nearest neighbors are attractive in the $z$-direction and repulsive in the $x$- and $y$-direction.
These attractive and repulsive magnetic interactions with correspondingly positive and negative second derivatives with respect to nearest-neighbor distances induce decrease and increase, respectively, of the Young moduli \cite{pessot2014structural}.
This trend is observed in Fig.~\ref{fig_cubic_EG_m} (a).
\begin{figure}[]
\centering
  \includegraphics[width=8.6cm]{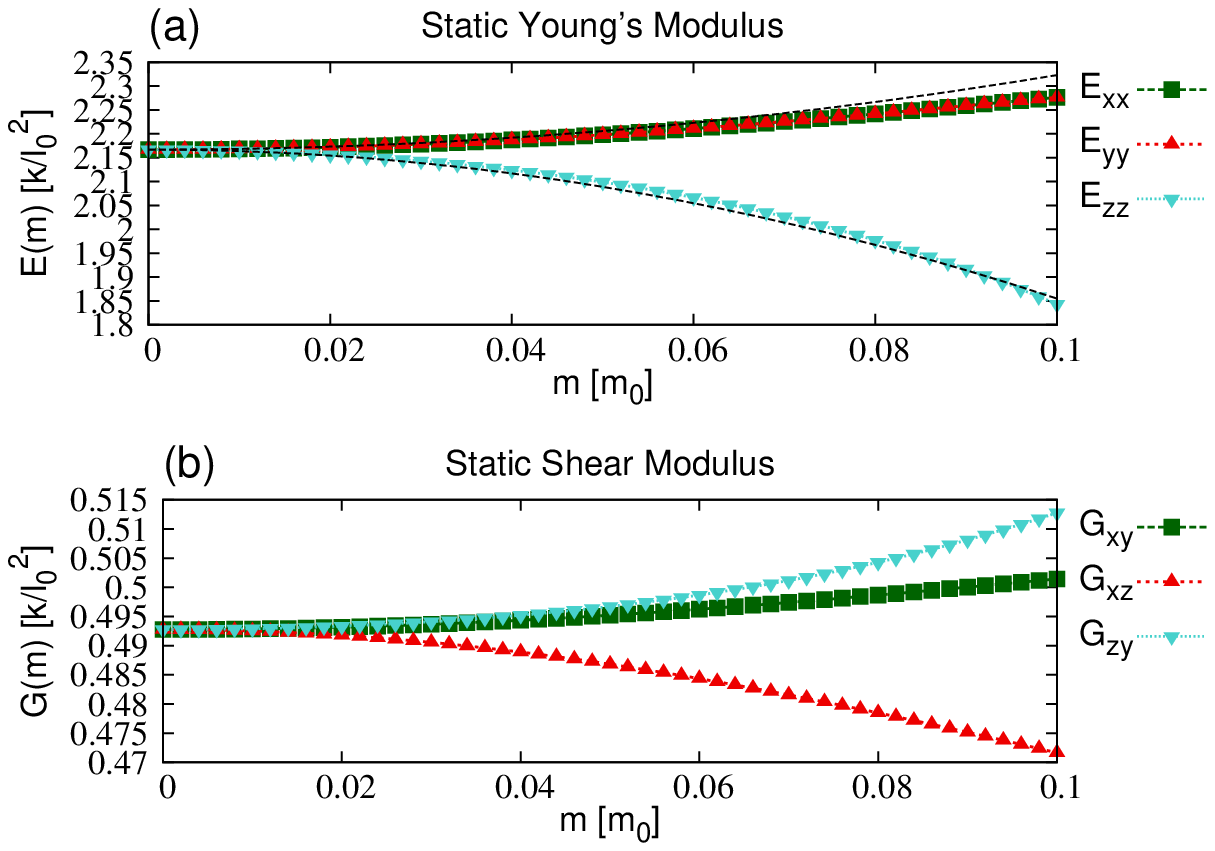}
  \includegraphics[width=8.6cm]{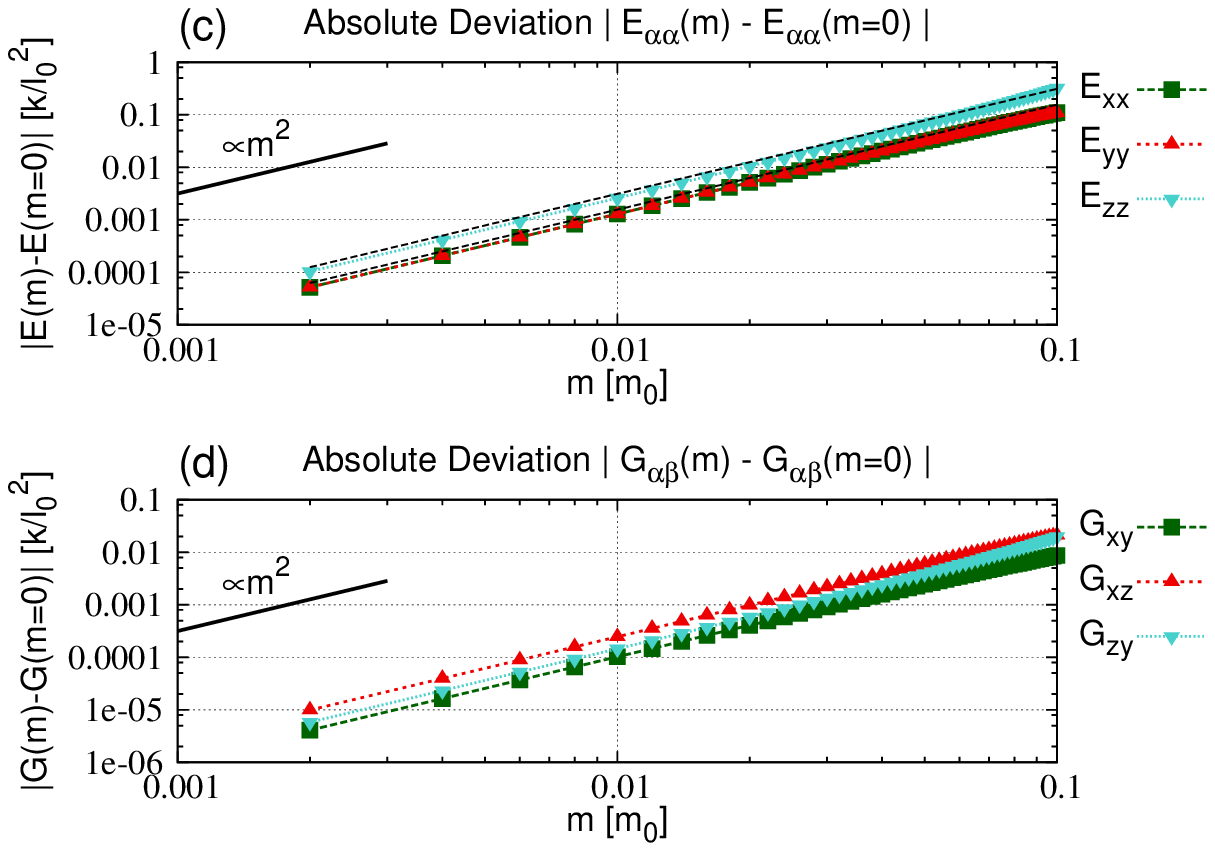}
  \caption{Static moduli \GP{(a)} $E_{\alpha\alpha}(m)=E'_{\alpha\alpha}(\omega=0,m)$ and \GP{(b)} $G_{\alpha\beta}(m)=G'_{\alpha\beta}(\omega=0,m)$ ($\alpha,\beta=x,y,z$) of a cubic lattice with $N=\npcube$ for increasing magnetic moment intensity $m$ ($\vm$ oriented along the $z$-direction).
(a) The Young moduli in the directions perpendicular to $\vm$ are increased by increasing magnetic moments, whereas in the $\vm$-direction the modulus is decreased.
Black dashed lines in panels (a) and (c) represent the trends in Eq.~(\ref{1n_sc_elm_m}) shifted vertically to compensate for finite-size and boundary effects and to allow for a better comparison of the $m$-dependence.
(b) The shear modulus $G_{xz}$ obtained by shear displacements along the $\vm$-direction decreases for increasing $m$, whereas $G_{xy}$ and $G_{zy}$ reveal an increasing behavior.
(c, d) All elastic moduli as functions of $m$ show quadratic behavior to lowest order, as required by the necessary $\vm \rightarrow - \vm$ symmetry.}
  \label{fig_cubic_EG_m}
\end{figure}
At vanishing magnetic moment all Young moduli measured along the different directions have the same value, as expected by the cubic lattice symmetry.
Then, as $m$ is slowly increased, this symmetry is broken and $E_{zz}(m)$ decreases, whereas $E_{xx}(m)$ and $E_{yy}(m)$ increase identically, as expected by the unbroken $x\leftrightarrow y$ symmetry.
Moreover, all moduli show to lowest order in $m$ a quadratic behavior, as demanded by the necessary $\vm \rightarrow - \vm$ symmetry \GP{\cite{ivaneyko2015dynamic}, see Fig.~\ref{fig_cubic_EG_m} (c).}

More explicitly, the trends of the static Young moduli in the simple cubic case can be explained by considering interactions between neighbors on a regular lattice, see appendix \ref{anyoung_app}.
When we focus on small magnetic interactions, i.e.\ $m\ll m_0$, the dipole--dipole forces are much weaker than the restoring elastic ones and we can assume they leave the particle positions unaltered.

Considering contributions up to neighbors as distant as $10\unitl$, we obtain, see appendix \ref{anyoung_app}, the following trends for the Young moduli
\begin{align}\label{1n_sc_elm_m}
\frac{E_{xx}(m)}{{k}\bigl/{\unitl^2}} = \frac{E_{yy}(m)}{{k}\bigl/{\unitl^2}}  &\approx \frac{9+4\sqrt{2}}{7} +15.61 {\left({m}/{m_0}\right)}^2, \nonumber \\
\frac{E_{zz}(m)}{{k}\bigl/{\unitl^2}} &\approx \frac{9+4\sqrt{2}}{7} -31.21 {\left({m}/{m_0}\right)}^2.
\end{align}

The trends provided by these expressions are in good agreement with our numerical results, see Fig.~\ref{fig_cubic_EG_m}~(a).
They describe, respectively, increasing or decreasing moduli in the directions perpendicular or parallel to $\vm$.
Moreover, Eq.~(\ref{1n_sc_elm_m}) suggests a stronger dependence of $E_{zz}$ on $m$ compared to $E_{xx}$ and $E_{yy}$.
This agrees with our numerical results, see Fig.~\ref{fig_cubic_EG_m} (a) and (c).
Furthermore, it confirms the major role played by the second derivatives of neighbor interactions in determining the trends for $E_{\alpha\alpha}(m)$ of regular distributions, as pointed out in Ref.~\cite{pessot2014structural}.

In our numerical calculations we obtain different behaviors for the different shear moduli as functions of $m$.
However, at vanishing magnetic moment they all assume the same value, as expected by lattice symmetry, see Fig.~\ref{fig_cubic_EG_m} (b).
Furthermore, as Young's moduli, they are all, to lowest order, quadratic functions of $m$, as required by symmetry when $\vm$ is flipped into $-\vm$, see Fig.~\ref{fig_cubic_EG_m} (d).
The shear modulus $G_{zy}(m)$ shows an increasing behavior for increasing $m$.
\GP{It is, in fact,} the only depicted shear deformation that breaks \GP{the spatial mutual} alignment of the moments in the $z$-direction.
This is hindered by increasing $m$, in agreement with an increasing modulus $G_{zy}(m)$.
The shear deformation related to $G_{xz}(m)$, instead, induces the dipoles to move in parallel to their alignment direction.
Nearest neighbors connected by $\unitl\dirx$ lie on a maximum of the dipole--dipole interaction, see Eq.~(\ref{emagn}).
Therefore, increasing $m$ facilitates the displacement induced by $\sigma_{xz}$, in agreement with a decreasing shear modulus $G_{xz}(m)$, as found in Fig.~\ref{fig_cubic_EG_m} (b).
Last, we find an increasing trend for the $G_{xy}(m)$ shear modulus, slightly weaker compared to the other two examined moduli, as depicted in Fig.~\ref{fig_cubic_EG_m} (b) and (d).

\subsection{Dynamic moduli}
We now focus on the dynamic properties, which are the central aim of the present work.
As a general trend, we always find the storage moduli to tend to a finite value for large $\omega$, see Fig.~\ref{fig_cubic_EG_wm}.
\begin{figure}[]
\centering
  \includegraphics[width=8.6cm]{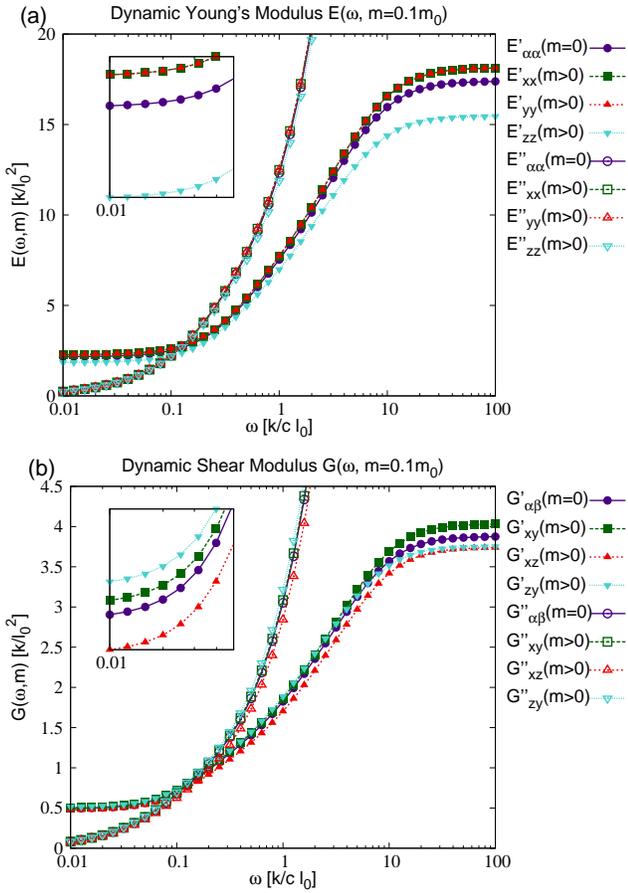}
  \caption{Dynamic elastic moduli (a) $E_{\alpha\alpha}(\omega)$ and (b) $G_{\alpha\beta}(\omega)$ ($\alpha,\beta=x,y,z$) of a cubic lattice with $N=\npcube$ for vanishing magnetic moment (solid line, $\bigcirc$), and $\vm=0.1 m_0 \dirz$ (dashed lines, $\Box,\bigtriangleup,\bigtriangledown$).
Filled and unfilled markers correspond to storage ($E'_{\alpha\alpha}$, $G'_{\alpha\beta}$) and loss ($E''_{\alpha\alpha}$, $G''_{\alpha\beta}$) components, respectively.
Insets in panels (a) and (b) zoom onto the storage parts (a) $E'_{\alpha\alpha}(\omega)$ and (b) $G'_{\alpha\beta}(\omega)$ at small $\omega$ for better resolution (see also Fig.~\ref{fig_cubic_EG_m}).}
  \label{fig_cubic_EG_wm}
\end{figure}
Yet, as noted before, it is not reasonable to consider the behavior for frequencies larger than $10 k/c\unitl$.
Conversely, the loss moduli as functions of $\omega$ show a linear increase (see appendix \ref{loss_app}).
This behavior we attribute to our model focusing on overdamped motion.
In fact, under oscillatory motion, the damping term in Eq.~(\ref{eq_motion}), which is the origin of the loss modulus, increases with frequency $\omega$.
This conforms with a macroscopic Kelvin-Voigt model \cite{mainardi2011creep,eldred1995kelvin} which predicts an imaginary component of the dynamic moduli linearly increasing with frequency.
\GP{Similarly, experimental measurements of the loss moduli in polymeric materials \cite{zanna2002influence,collin2003frozen,roeben2014magnetic,hohlbein2015remote} are compatible with a Kelvin-Voigt model [i.e.\ constant storage part and linearly increasing loss part of $E_{\alpha\alpha}(\omega)$ and $G_{\alpha\beta}(\omega)$] in the low-frequency regime.}
Furthermore, in the limit $\omega\rightarrow 0$, we always find vanishing loss moduli and the storage component to recover the corresponding static elastic modulus, see Eqs.~(\ref{stat_elmod}), (\ref{kappa}), and (\ref{dyn_elm}).

The storage Young moduli $E'_{\alpha\alpha}(\omega)$ ($\alpha=x,y,z$) in Fig.~\ref{fig_cubic_EG_wm} (a)---here calculated for $\vm=0.1m_0\dirz$---show at all frequencies the trends as described in the static case, see Fig.~\ref{fig_cubic_EG_m}.
The amount of variation with respect to the $m=0$ configuration, however, seems to be larger at larger frequencies.
Furthermore, $E'_{xx}(\omega)$ and $E'_{yy}(\omega)$ show identical behavior as functions of $\omega$, as required by the symmetry of this geometry under switching $x\leftrightarrow y$.
Likewise, at low $\omega$, the loss moduli $E''_{zz}(\omega)$ and $E''_{xx,yy}(\omega)$ show a decreasing and increasing trend, respectively, when the magnetic moment is switched on and increased.
Furthermore, for higher $\omega$, all the loss components linearly increase with $\omega$ with identical coefficients, \GP{see also Fig.~\ref{fig_app_loss_cubic} (b) in appendix \ref{loss_app}.}

The storage shear moduli $G'_{\alpha\beta}(\omega)$ at low frequencies present the same trends of increase and decrease as in the static case, see Figs.~\ref{fig_cubic_EG_m} (b) and \ref{fig_cubic_EG_wm} (b).
We remark that at high frequencies (beyond $10 k/c\unitl$), while $G'_{xy}(\omega)$ and $G'_{xz}(\omega)$ show the same and enhanced trend as in the static case, $G'_{zy}(\omega)$ changes from increase to decrease by increasing $m$.
This graphically results in a crossing between the curves for $G'_{\alpha\beta}(\omega,m=0)$ and $G'_{zy}(\omega,m=0.1m_0)$.
The loss shear moduli $G''_{\alpha\beta}(\omega)$, instead, display the same increasing or decreasing trends as the corresponding static $G_{\alpha\beta}(m=0)$ both at low and high frequencies (see also appendix \ref{loss_app}).

\section{Fcc Lattice}\label{fcc_result}
We now turn our focus onto the exemplary case of a face-centered cubic (\fcc) lattice.
Later in section \ref{disord_result}, we will generate disordered samples by randomizing an initially \fcc\ particle arrangement.
In this setup we introduce springs connecting nearest neighbors only.
This is enough to obtain a particle distribution stable under both stretching and shearing.
The boundaries of the system are chosen as the outermost layers of particles in a given direction.
The typical interparticle distance $\unitl$ follows from the number density $\rho$, as explained in section \ref{dip_spring_model}, which for the \fcc\ lattice is $4$ particles per unit cell.

When magnetic moments are introduced we here observe an elongation of the system in the $\vm$-direction and shrinking in the perpendicular directions, see Fig.~\ref{fig_fcc_latt}.
\begin{figure}[]
\centering
  \includegraphics[width=8.6cm]{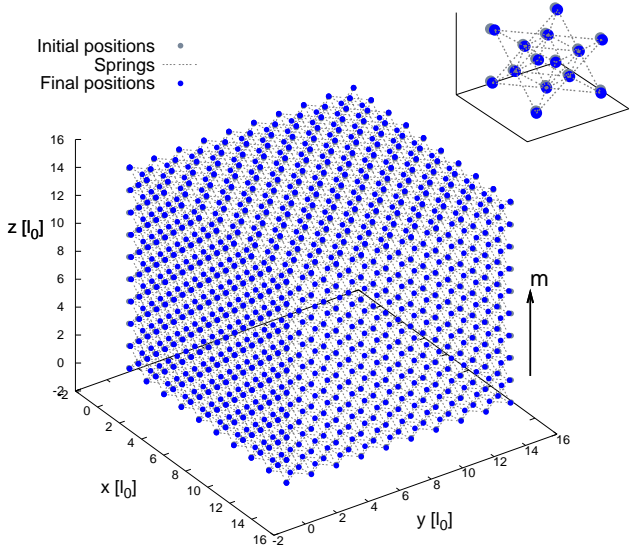}
  \caption{Deformation of an \fcc\ lattice with springs between nearest neighbors and $N=\npfcc$, when a magnetic moment of $\vm=0.1 m_0\dirz$ is switched on.
For illustrative purposes, only the first two particle layers on the front, top, and right faces are depicted.
Elongation is observed in the $\vm$-direction and contraction in the perpendicular ones.
Inset zooms in onto the deformation of the particles at the bottom left corner of the sample.}
  \label{fig_fcc_latt}
\end{figure}
The nearest neighbors on the \fcc\ lattice are located along the $\dirx+\diry$, $\dirx+\dirz$, and $\diry+\dirz$ directions, i.e.\ at an angle of $\pi/4$ with respect to the Cartesian axes.
When the system elongates in the $z$-direction the angles between the nearest-neighbor directions and $\vm$ reduce, thus lowering the magnetic energy $U^m$.

\subsection{Static moduli}\label{fcc_result_static}
First, we present the behavior of the static moduli as functions of increasing magnetic moment, see also Ref.~\cite{ivaneyko2012effects}.
We always find a monotonic, smooth behavior for increasing $m$ [see Fig.~\ref{fig_fcc_EG_m} \GP{(a) and (b)}].
\begin{figure}[]
\centering
  \includegraphics[width=8.6cm]{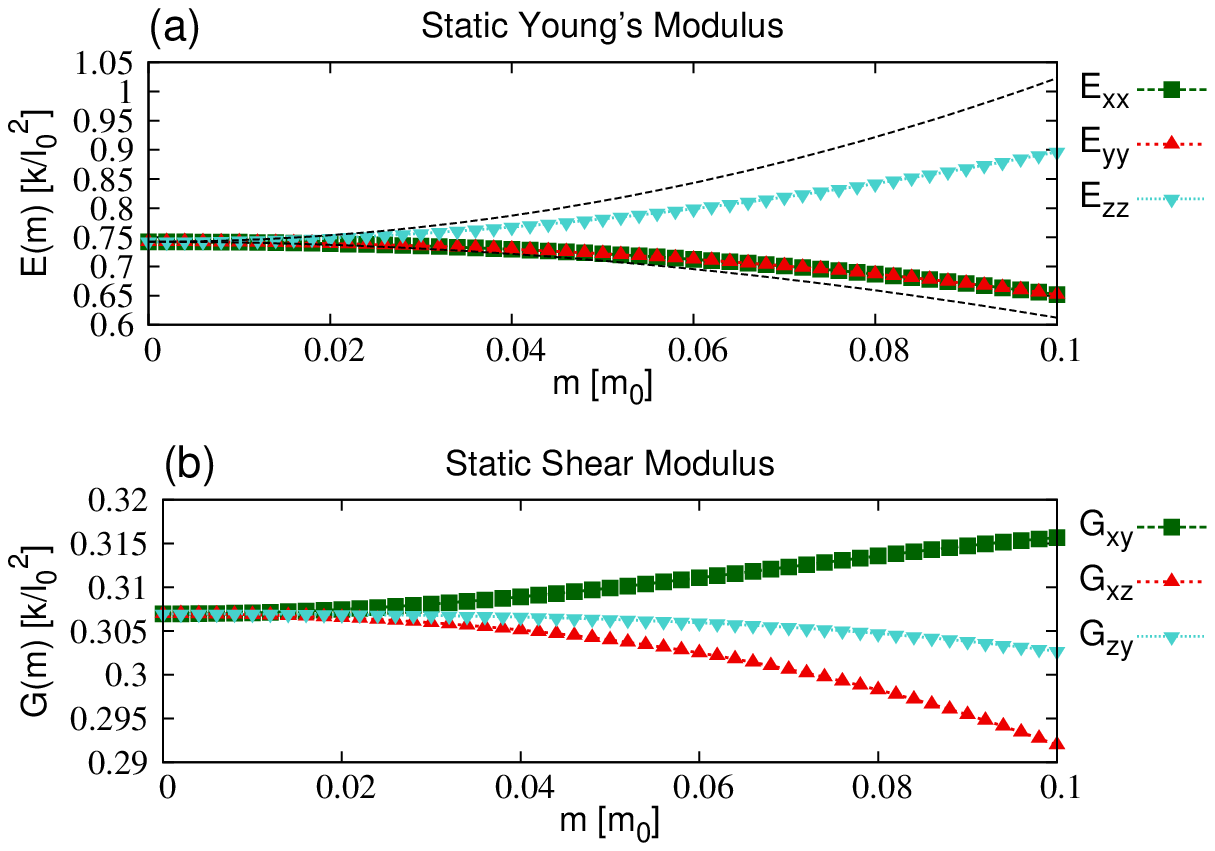}
  \includegraphics[width=8.6cm]{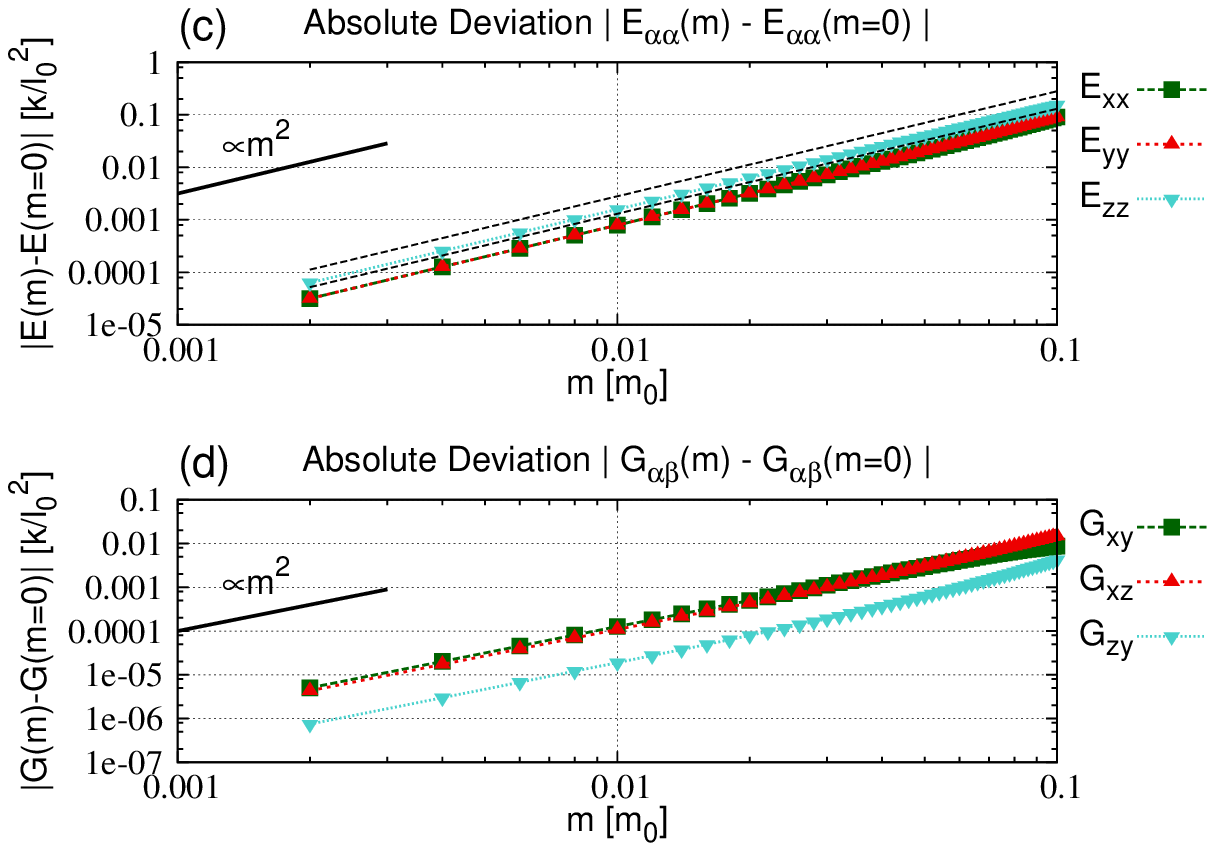}
  \caption{Static moduli (a) $E_{\alpha\alpha}(m)=E'_{\alpha\alpha}(\omega=0,m)$ and (b) $G_{\alpha\beta}(m)=G'_{\alpha\beta}(\omega=0,m)$ ($\alpha,\beta=x,y,z$) of an \fcc\ lattice with $N=\npfcc$ for increasing magnetic moment intensity $m$.
$\vm$ is oriented along the $z$-direction.
The Young moduli for stretching perpendicular to $\vm$ are reduced by increasing magnetic moments, whereas along the $\vm$-direction the modulus is increased.
Black dashed lines in panels (a) and (c) represent the trends in Eq.~(\ref{1n_fcc_elm_m}) shifted vertically to compensate for finite-size and boundary effects and to allow for a better comparison of the $m$-dependence.
The shear modulus $G_{xy}$ obtained by applying shear in the $xy$ plane perpendicular to $\vm$ increases for increasing $m$, whereas $G_{xz}$ and $G_{zy}$ reveal a decreasing behavior.
\GP{(c, d)} All elastic moduli as functions of $m$ show a quadratic behavior to lowest order, in accord with the $\vm \rightarrow - \vm$ symmetry and as depicted by the log-log scale plots.}
  \label{fig_fcc_EG_m}
\end{figure}
In fact, as shown in Fig.~\ref{fig_fcc_EG_m} \GP{(c) and (d)}, the elastic moduli as functions of $m$ are to lowest order quadratic functions, in accord with the $\vm \rightarrow - \vm$ symmetry.
Again, and as required by lattice symmetry, at $m=0$ all Young moduli and the shear moduli in the examined directions coincide, see Fig.~\ref{fig_fcc_EG_m} \GP{(a) and (b)}.

Next, we estimate  the role played by the relative positions of neighboring particles for the behavior of the Young moduli.
We consider the case of a regular \fcc\ lattice and take into account contributions to the Young moduli to lowest order in $m$, as explained in appendix \ref{anyoung_app}.
Considering terms up to neighbors as far as $10\unitl$ in Eq.~(\ref{an_stiff_mod}), we obtain
\begin{align}\label{1n_fcc_elm_m}
\frac{E_{xx}(m)}{{k}\bigl/{\unitl^2}}=\frac{E_{yy}(m)}{{k}\bigl/{\unitl^2}}  &\approx \frac{2^{7/6}}{3} -13.02 {\left({m}/{m_0}\right)}^2, \nonumber \\
\frac{E_{zz}(m)}{{k}\bigl/{\unitl^2}} &\approx \frac{2^{7/6}}{3} +28.05 {\left({m}/{m_0}\right)}^2.
\end{align}
Comparison with the behavior of the Young's moduli resulting from our numerical calculations, see Fig.~\ref{fig_fcc_EG_m} (a), leads to a good qualitative agreement.
The modulus in the $\vm$-direction $E_{zz}(m)$ increases with increasing $m$, whereas in the perpendicular directions $E_{xx}(m)$ and $E_{yy}(m)$ decrease with $m$.
Thus, the \fcc\ arrangement shows a completely opposite behavior compared to the simple cubic case, see section \ref{sc_result_static}.
Moreover, Eq.~(\ref{1n_fcc_elm_m}) indicates the $E_{zz}(m)$ modulus to have a stronger dependence on $m$ compared to $E_{xx}(m)$ and $E_{yy}(m)$, as also found in our numerical results and shown in Fig.~\ref{fig_fcc_EG_m} \GP{(a) and (c)}.

Similarly, the shear moduli are influenced by $m$ in different ways.
Here we find the shear modulus $G_{xy}(m)$ to increase and $G_{xz}(m)$ to decrease with increasing $m$, analogously to what we observed in the simple cubic case, see section \ref{sc_result_static}.
Contrarily to the simple cubic case, the shear modulus referring to displacements parallel to $\vm$, $G_{zy}(m)$, shows a decreasing trend when the magnetic moments increase.
Moreover, $G_{zy}(m)$ displays a weaker dependence on $m$ compared to the remaining two shear moduli, as depicted in Fig.~\ref{fig_fcc_EG_m} (d).

\subsection{Dynamic moduli}
Finally, we examine the behaviors of the dynamic elastic moduli for various frequencies $\omega$ and magnetic moment intensities $m$.
The storage dynamic Young moduli $E'_{\alpha \alpha}$ ($\alpha=x,y,z$) at all frequencies follow the same behavior as described in the static case (see Fig.~\ref{fig_fcc_EG_m}).
In the direction parallel to $\vm$, $E'_{zz}$ increases for increasing $m$, whereas $E'_{xx}$ and $E'_{yy}$ decrease for the perpendicular directions, see Fig.~\ref{fig_fcc_EG_wm} (a) and its inset for a zoom onto the low-$\omega$ behavior.
\begin{figure}[]
\centering
  \includegraphics[width=8.6cm]{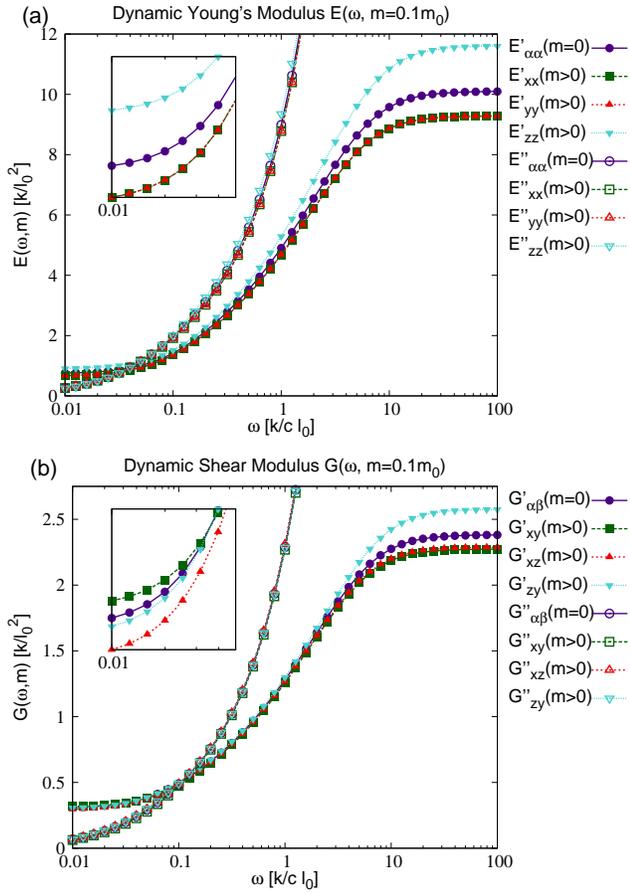}
  \caption{Dynamic elastic moduli (a) ${E}_{\alpha\alpha}(\omega)$ and (b) ${G}_{\alpha\beta}(\omega)$ ($\alpha,\beta=x,y,z$) of an \fcc\ lattice with $N=\npfcc$ for vanishing magnetic moment (solid line, $\bigcirc$), and $\vm=0.1 m_0 \dirz$ (dashed lines, $\Box,\bigtriangleup,\bigtriangledown$).
Filled and unfilled markers correspond to storage ($E'$, $G'$) and loss ($E''$, $G''$) components, respectively.
Insets in panels (a) and (b) zoom onto the storage parts (a) $E'_{\alpha\alpha}(\omega)$ and (b) $G'_{\alpha\beta}(\omega)$ at small $\omega$ to better resolve the different curves (see also Fig.~\ref{fig_fcc_EG_m}).}
  \label{fig_fcc_EG_wm}
\end{figure}
As shown in appendix \ref{loss_app}, the loss components $E''_{\alpha \alpha}$ partially exhibit opposite trends compared to their storage counterparts (see Fig.~\ref{fig_app_loss_fcc} for a detailed plot).
In fact, at low frequencies, the loss modulus for the $\vm$ direction, $E''_{zz}$, decreases with increasing $m$, whereas for $E''_{xx}$ and  $E''_{yy}$ the two perpendicular directions increase.
At higher frequencies, however, and as in the cubic lattice case, all the loss moduli $E''_{\alpha \alpha}$ recover the behavior of their storage counterparts and show an identical dependence on $\omega$ [see Figs.~\ref{fig_fcc_EG_wm} (a) and Fig.~\ref{fig_app_loss_fcc} in appendix \ref{loss_app}].

The storage dynamic shear moduli $G'_{\alpha \beta}$ ($\alpha, \beta=x,y,z$) are displayed in Fig.~\ref{fig_fcc_EG_wm} \GP{(b)}.
Here, at low-$\omega$ values the changes in the shear moduli for the different geometries reproduce the trends shown in Fig.~\ref{fig_fcc_EG_m}, see the inset of Fig.~\ref{fig_fcc_EG_wm} (b).
However, when considering the behavior at higher $\omega$, $G'_{xy}$ turns from increasing to decreasing with $m$, while $G'_{zy}$ turns from decreasing to increasing when compared with the shear modulus at $m=0$.
Although we already mentioned that only the behavior for $\omega \lesssim 10 k/c\unitl$ should be interpreted, these data suggest the possibility that some dynamic shear moduli could swap their tendency of increasing or decreasing with $m$ to decreasing or increasing, respectively.
Contrarily, the Young moduli consistently show a monotonic behavior as functions of both $\omega$ and $m$.
Furthermore, at low $\omega$, the loss shear moduli $G''_{\alpha \beta}$ exhibit an opposite behavior when compared with their storage complements.
For shear deformations in the plane perpendicular to $\vm$, $G''_{xy}$ decreases with increasing magnetic moment, whereas the other two moduli $G''_{xz}$ and $G''_{zy}$ are increased by increasing $m$, see also appendix \ref{loss_app}, Fig.~\ref{fig_app_loss_fcc}.

\section{$3$D disordered samples}\label{disord_result}
\subsection{Numerical generation}\label{disord_generation}
We start from a regular three-dimensional \fcc\ lattice.
Having a well defined density $\rho$ and neighbor structure, this lattice allows us to define the average interparticle distance $\unitl$ as described in section \ref{fcc_result}.
Then we introduce disorder in the lattice by randomly displacing each particle by $0.5 \unitl$ in a stochastic direction.
After that, we set the elastic springs between nearest neighbors.

In the randomization step, we take care to generate an initially stable disordered system so that magnetic interactions do not immediately overcome the elastic spring interactions when the magnetic moments are switched on \cite{biller2014modeling,annunziata2013hardening}.
In other words, the formation of collapsed clusters where the particles touch each other in a stuck configuration shall be avoided for low strength of the magnetic interactions.
For this purpose, we impose that in the randomized configuration for $m=0$ no couples of particles are closer than $0.5 \unitl$.
Boundary particles are identified as the outermost layers of the initial \fcc\ lattice in each direction.
To help maintain an overall cubelike shape, we move boundary particles by half the amount of other particles.
An example of the resulting initial distribution is given by the gray particles in Fig.~\ref{fig_dis_fcc_lattice}.
\begin{figure}[]
\centering
  \includegraphics[width=8.6cm]{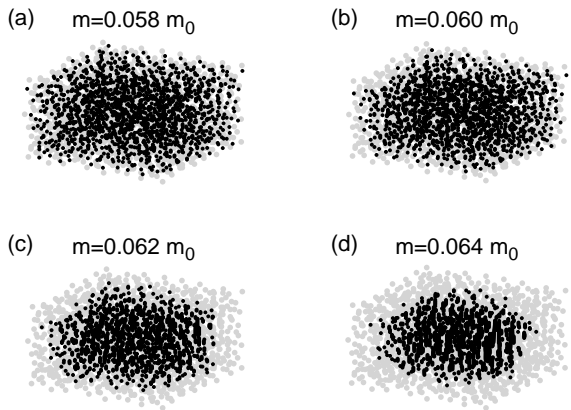}
  \caption{Example deformation of a randomized particle distribution (N=\npdisord) of initially cubelike shape (gray particles) when a magnetic moment of $\vm=m \dirz$ is switched on.
Panels (a), (b), (c), and (d) show the equilibrated particle distribution (black) as the magnetic moment intensity is gradually increased to $m=0.058 m_0$, $m=0.06 m_0$, $m=0.062 m_0$, and $m=0.064 m_0$, respectively.
Panel (c) represents the onset of chain formation in the $\vm$-direction, see sections \ref{dis_fcc_stat_m} and \ref{dis_fcc_dyn_highm}.}
  \label{fig_dis_fcc_lattice}
\end{figure}

Thus, we generate a disordered system of macroscopic cubelike shape with $N$ non-overlapping magnetic particles.
In the following we set $N=\npdisord$.
As described, in the initial configuration, the springs are set before the magnetic interactions are switched on.
Then, we gradually increase the magnitude of the magnetic moments and at each step find the minimum energy configuration, see section \ref{equil_state}.
When the equilibrium state for a given $m$ is reached, we obtain the Young and shear moduli $E$ and $G$ as functions of both $m$ and $\omega$, using the methods described in sections \ref{stat_elm_calc} and \ref{dynamic_elm}.

As the magnitude $m$ of the magnetic moments increases, we can principally distinguish between two regimes.
On the one hand, the behavior for small $m$ is controlled by magnetic $U^m$ and elastic $U^{el}$ energies, see Fig.~\ref{fig_dis_fcc_GS_en}.
\begin{figure}[]
\centering
  \includegraphics[width=8.6cm]{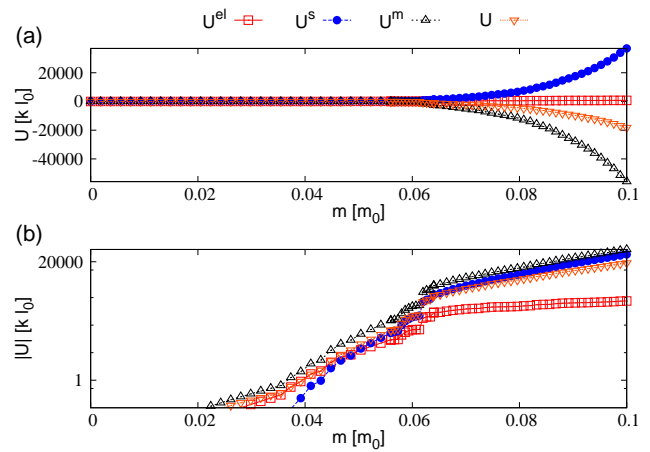}
  \caption{(a) Equilibrium energies of the disordered \fcc\ system shown in Fig.~\ref{fig_dis_fcc_lattice} for increasing magnitude of the magnetic moment $m$.
\GP{(b) Two regimes are identified in a logarithmic plot.
Up to $m\sim 0.05 m_0$ the total energy $U$ mostly comprises elastic $U^{el}$ and magnetic $U^{m}$ contributions.
For $m\gtrsim 0.05 m_0$ instead, the steric interaction energy $U^s$ becomes higher than the elastic energy $U^{el}$.
This signals the subsequent formation of chains.
The pronounced step at $0.06 m_0 \lesssim m \lesssim 0.064 m_0$ is connected to chain formation.}
}
  \label{fig_dis_fcc_GS_en}
\end{figure}
The deformation is relatively small and the elastic moduli are, to lowest order, quadratic functions of $m$, as expected by the necessary $\vm \rightarrow - \vm$ symmetry.
On the other hand, when attractive magnetic interactions become as strong as to overcome linear spring repulsion, steric interactions come into play (see Fig.~\ref{fig_dis_fcc_GS_en}).
Then, formation of chains is observed, as well as significant changes in the system size (see Fig.~\ref{fig_dis_fcc_lattice}).
Furthermore, the close steric contact between particles generates extra stiffness, which is reflected by a significant change in the elastic moduli.
This behavior reflects a ``hardening transition'' similar to the situation described in Ref.~\cite{annunziata2013hardening} for one-dimensional systems.

\subsection{Static moduli}\label{dis_fcc_stat_m}
First, we focus on the static elastic moduli of the randomized system for increasing magnetic moment $m$.
To extract a general trend we realized $\ndissamples$ different systems following the protocol as described in section \ref{disord_generation}.
Then we obtain our results by averaging over the moduli for all different randomized realizations.
Relative errors follow from the standard deviations.
The resulting static moduli are depicted in Fig.~\ref{fig_dis_fcc_EG_m}.
\begin{figure}[]
\centering
  \includegraphics[width=8.6cm]{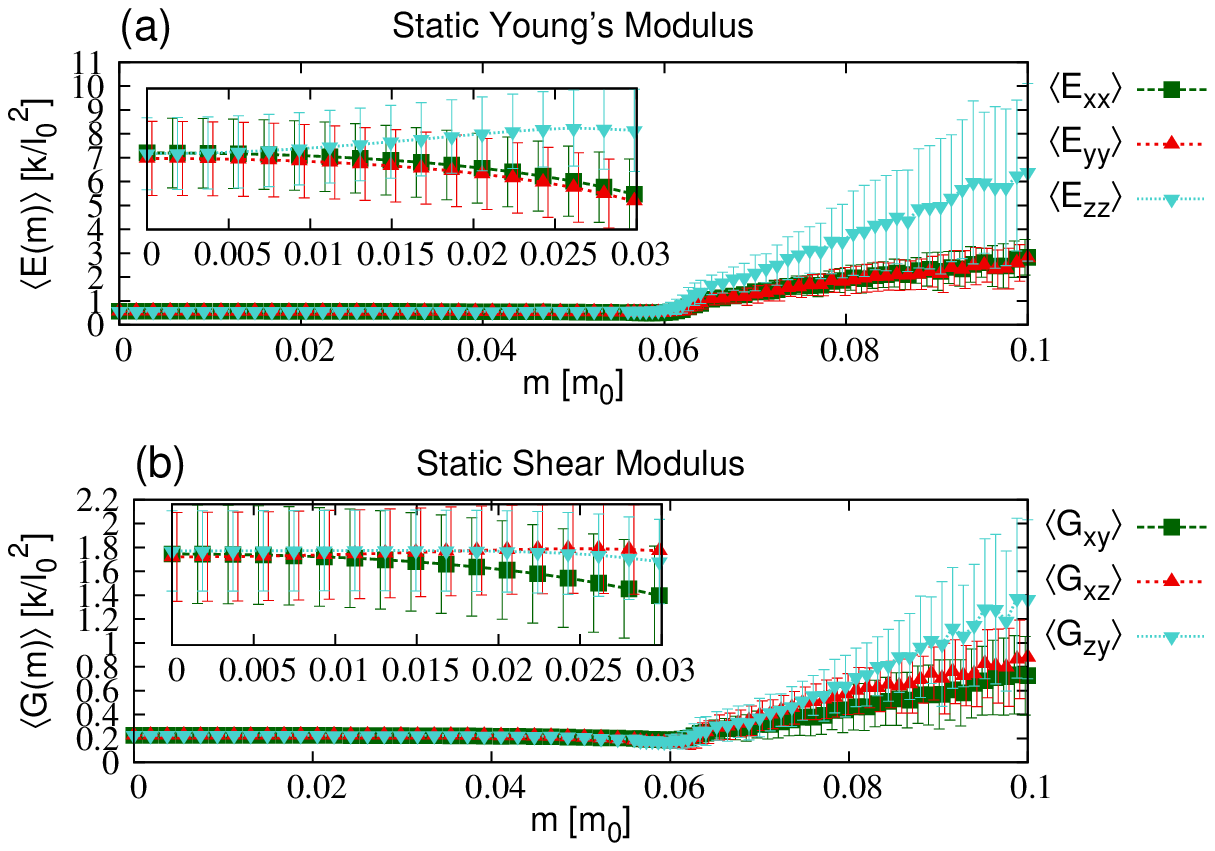}
  \includegraphics[width=8.6cm]{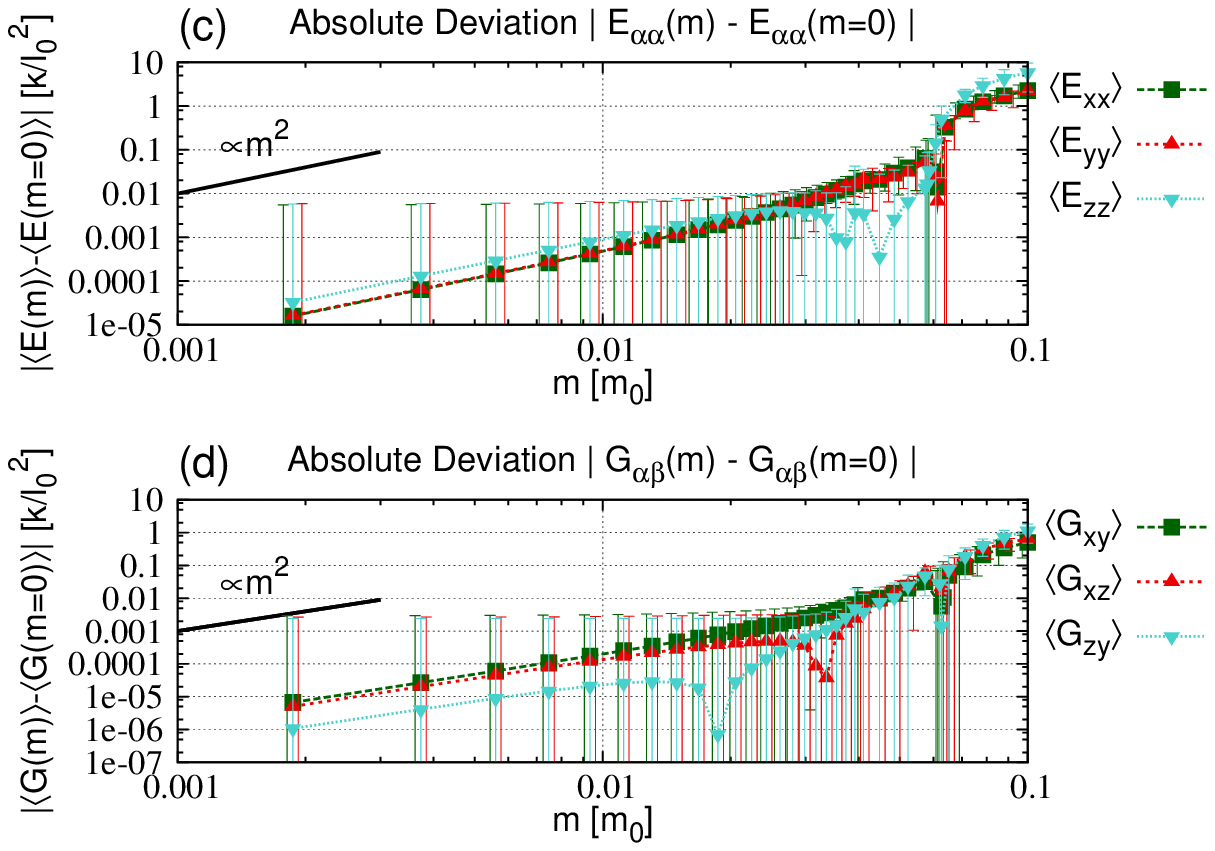}
  \caption{Static moduli (a) $\langle E_{\alpha\alpha}(m) \rangle=\langle E'_{\alpha\alpha}(\omega=0,m)\rangle$ and (b) $\langle G_{\alpha\beta}(m)\rangle=\langle G'_{\alpha\beta}(\omega=0,m)\rangle$ ($\alpha,\beta=x,y,z$) of a disordered \fcc\ lattice with $N=\npdisord$ for increasing $m=|\vm|$, with $\vm$ oriented in the $z$-direction.
Statistics are collected over $\ndissamples$ differently randomized samples.
Data points and bars represent the resulting averages and standard deviations, respectively.
(c, d) All elastic moduli as functions of $m$ show a quadratic behavior to lowest order for small $m$, in accord with the $\vm \rightarrow - \vm$ symmetry.
\GP{For illustrative purposes we have slightly shifted the bars for different data sets horizontally and reduced the number of points shown in panels (c) and (d) to better distinguish between the individual bars and data points.}
\GP{Dips in panels (c) and (d) occur when $\langle E_{\alpha\alpha}(m) \rangle\approx\langle E_{\alpha\alpha}(m=0) \rangle$ or $\langle G_{\alpha\beta}(m) \rangle\approx\langle G_{\alpha\beta}(m=0) \rangle$.
Then, the logarithm of the absolute deviation from the value for $m=0$ diverges to $-\infty$.
The elastic moduli themselves, however, show smooth behavior, as displayed in panels (a) and (b) and respective insets.}}
  \label{fig_dis_fcc_EG_m}
\end{figure}
To lowest order in $m$ and up to approximately $m=0.06 m_0$, the Young moduli of the system [see inset of Fig.~\ref{fig_dis_fcc_EG_m} (a)] show a behavior similar to the \fcc\ case [compare with Fig.~\ref{fig_fcc_EG_m} (a)]: increasing $\langle E_{zz} \rangle$ for imposed deformations in the $\vm$ direction and decreasing $\langle E_{xx} \rangle$ and $\langle E_{yy} \rangle$ for the perpendicular cases.
Moreover, in this regime the static Young moduli $\langle E_{\alpha\alpha}(m)\rangle$ ($\alpha=x,y,z$) show a quadratic behavior as functions of $m$ in accord with the $\vm \rightarrow - \vm$ symmetry, see Fig.~\ref{fig_dis_fcc_EG_m} (c).
Similarly, the static shear moduli $\langle G_{\alpha\beta}(m)\rangle$ ($\alpha,\beta=x,y,z$) in this regime show quadratic behavior, see Fig.~\ref{fig_dis_fcc_EG_m} (d), while the trends for $\langle G_{\alpha\beta}(m)\rangle$ vary from those of the regular \fcc\ lattice [compare the inset of Fig.~\ref{fig_dis_fcc_EG_m} (b) with Fig.~\ref{fig_fcc_EG_m} (b)].

This behavior changes dramatically for $m \gtrsim 0.06 m_0$, where magnetic interactions are as strong as to cause the particles to come into steric contact and form chains in the $\vm$-direction.
Here we observe a significant increase in all elastic moduli [see Fig.~\ref{fig_dis_fcc_EG_m} (a) and (b)].
Still, Young's modulus for imposed deformations in the $\vm$-direction, $\langle E_{zz}\rangle$, shows a much larger increase compared to $\langle E_{xx}\rangle$ and $\langle E_{yy}\rangle$, in agreement with experimental observations \GP{on} anisotropic systems \cite{varga2006magnetic}, see also the case of bi-axial tension \cite{schubert2016equi}.
$\langle E_{xx}\rangle$ and $\langle E_{yy}\rangle$ feature an identical behavior within the errorbars, as expected by the largely unbroken isotropy of the systems within the $xy$-plane.
Likewise, the shear moduli show an increase for all investigated geometries.
In a purely affine deformation of chains perfectly aligned along $\vm$, the $zy$ shear geometry would be the only one displayed that leads to distortions of the chains.
Therefore it is conceivable that $\langle G_{zy}\rangle$ grows larger than $\langle G_{xy}\rangle$ and $\langle G_{xz}\rangle$, although the size of the \GP{standard deviations} does not allow to draw a conclusive result.

\GP{Finally, to avoid confusion, we stress that the dips in Fig.~\ref{fig_dis_fcc_EG_m} (c) and (d) simply mean that the elastic moduli for $m\neq 0$ tend to the same values as those for $m=0$.
Since in Fig.~\ref{fig_dis_fcc_EG_m} (c) and (d) the deviations of the elastic moduli from their values for $m=0$ are plotted on a logarithmic scale, the dips are not directly related to a mechanical instability resulting from vanishing elastic moduli.
In fact, as shown in in Fig.~\ref{fig_dis_fcc_EG_m} (a) and (b), for a given value of $m$ the elastic moduli always remain positive.}

\subsection{Dynamic moduli, $m\lesssim 0.06 m_0$}\label{dis_fcc_dyn_lowm}
We now move our attention to the dynamic properties of our disordered systems.
Again, we have collected statistics over $\ndissamples$ different realizations of our randomizing process.
The resulting averages and standard deviations are represented as data points and bars in the figures below.

First we examine the dynamic moduli for the magnitude of the magnetic moments below the onset of significant chain formation, i.e.\ $m\lesssim 0.06 m_0$.
There, the storage parts $\langle E'_{\alpha\alpha}(\omega)\rangle$ of the dynamic Young moduli for increasing $m$ show the same trends for the different geometries as the static moduli [see Fig.~\ref{fig_dis_fcc_EG_wm_low} (a) and compare with the inset of Fig.~\ref{fig_dis_fcc_EG_m} (a)].
\begin{figure}[]
\centering
  \includegraphics[width=8.6cm]{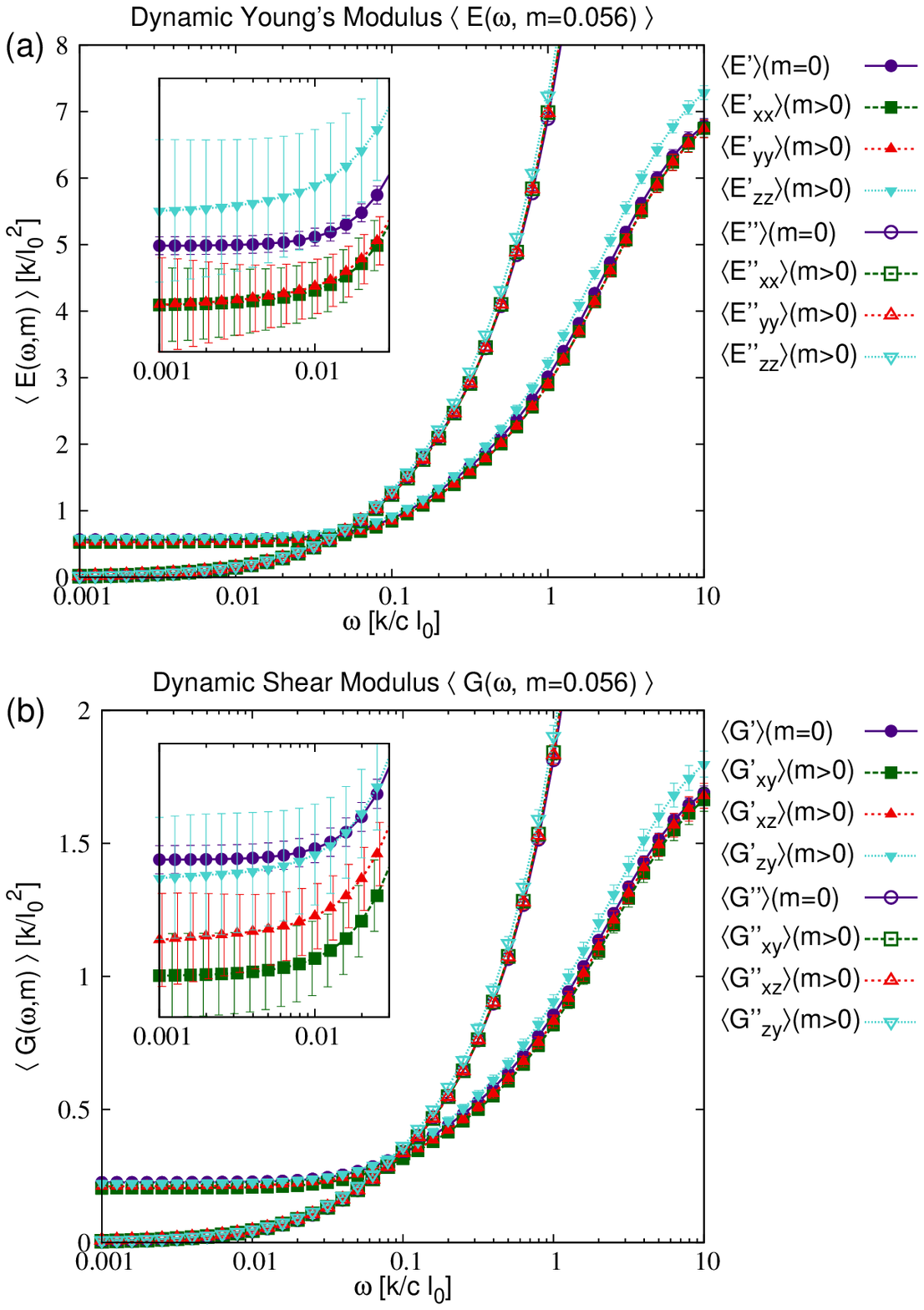}
  \caption{Dynamic elastic moduli (a) $\langle E_{\alpha\alpha}(\omega)\rangle$ and (b) $\langle G_{\alpha\beta}(\omega)\rangle$ ($\alpha,\beta=x,y,z$) of randomized \fcc\ lattices with $N=\npdisord$ for vanishing magnetic moment (solid line, $\bigcirc$), and $\vm=\mlow m_0 \dirz$ (dashed lines, $\Box,\bigtriangleup,\bigtriangledown$).
Data points and standard deviations are obtained by averaging over $\ndissamples$ differently randomized samples.
Filled and unfilled markers correspond to storage ($E'$, $G'$) and loss ($E''$, $G''$) components, respectively.
Insets zoom onto the storage parts (a) $\langle E'_{\alpha\alpha}(\omega)\rangle$ and (b) $\langle G'_{\alpha\beta}(\omega)\rangle$ at small $\omega$ to better resolve the different curves.
\GP{For illustrative purposes we have slightly shifted the bars for different data sets horizontally to better distinguish the individual bars.}}
  \label{fig_dis_fcc_EG_wm_low}
\end{figure}
Conversely, the loss parts $\langle E''_{\alpha\alpha}(\omega)\rangle$ of the Young moduli feature a trend of increase with increasing $m$ in all cases [see appendix \ref{loss_app}, Fig.~\ref{fig_app_loss_dis_fcc_low} (a)].

Similarly to the Young moduli, the storage parts $\langle G'_{\alpha\beta}(\omega)\rangle$ of the dynamic shear moduli approximately follow their static counterparts at low $\omega$ [see the inset of Fig.~\ref{fig_dis_fcc_EG_wm_low} (b) and compare it to the inset of Fig.~\ref{fig_dis_fcc_EG_m} (b)].
However, with increasing frequencies $\omega$ and upon switching $m$ from $m=0$ to $m>0$, $\langle G'_{zy}(\omega)\rangle$ switches from a slight decrease to a significant increase with respect to the value at $m=0$, see Fig.~\ref{fig_dis_fcc_EG_wm_low} (b).
This results in a crossing between the curves corresponding to $\langle G'_{zy}(\omega,m=0)\rangle$ and $\langle G'_{zy}(\omega,m>0)\rangle$.
Instead, the remaining two shear moduli $\langle G'_{xy}(\omega)\rangle$ and $\langle G'_{xz}(\omega)\rangle$ always show a decrease.
Analogously to $\langle E''_{\alpha\alpha}(\omega)\rangle$, the loss components $\langle G''_{\alpha\beta}(\omega)\rangle$ are observed to increase at all frequencies when switching on $m$, independently of the chosen geometry [see appendix \ref{loss_app}, Fig.~\ref{fig_app_loss_dis_fcc_low} (b)].

\subsection{Dynamic moduli, $m \gtrsim 0.06 m_0$}\label{dis_fcc_dyn_highm}
In the following, we consider the dynamic moduli of the system at magnitudes $m$ of the magnetic moment at the onset of chain formation [see Fig.~\ref{fig_dis_fcc_lattice} (c)].
Then steric interactions play a major role in the total interaction energy $U$ (see Fig.~\ref{fig_dis_fcc_GS_en}).
\GP{To better illustrate the behavior of the storage dynamic moduli in this regime it is convenient to plot the \textit{deviation} from the respective static value at $m=0$, as shown in Fig.~\ref{fig_dis_fcc_EG_wm_high} (for brevity, although deviations are plotted, the curves are still labeled by $\langle E'_{\alpha\alpha}\rangle$ and $\langle G'_{\alpha\beta}\rangle$).}
\begin{figure}[]
\centering
  \includegraphics[width=8.6cm]{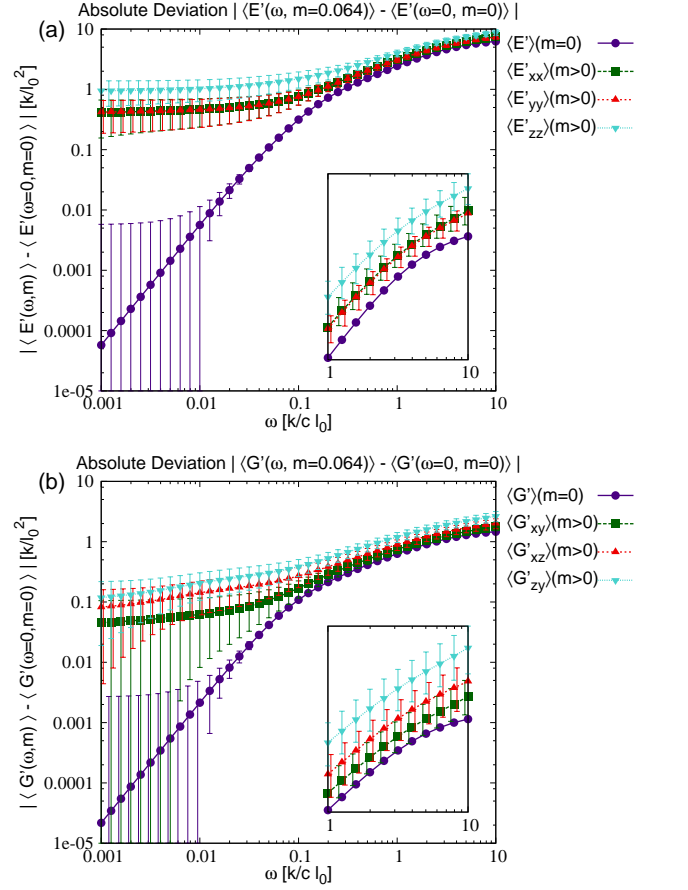}
  \caption{Storage dynamic elastic moduli (a) $\langle E'_{\alpha\alpha}(\omega)\rangle$ and (b) $\langle G'_{\alpha\beta}(\omega)\rangle$ ($\alpha,\beta=x,y,z$) of randomized \fcc\ lattices with $N=\npdisord$ for vanishing magnetic moment (solid line, $\bigcirc$), and $\vm=\mhigh m_0 \dirz$ (dashed lines, $\Box,\bigtriangleup,\bigtriangledown$).
We plot on a double logarithmic scale the absolute deviation \GP{from} the respective average static modulus at $m=0$.
Data points and standard deviations are obtained from statistics over $\ndissamples$ differently randomized samples.
\GP{For illustrative purposes we have slightly shifted the bars for different data sets horizontally to better distinguish the individual bars.}
Insets zoom onto the storage parts (a) $\langle E'_{\alpha\alpha}(\omega)\rangle$ and (b) $\langle G'_{\alpha\beta}(\omega)\rangle$ at large $\omega$ to better resolve the different curves.
\GP{Small values of the curves for the $m=0$ cases at low $\omega$ indicate smooth convergence to the respective static moduli in Fig.~\ref{fig_dis_fcc_EG_m}.}}
  \label{fig_dis_fcc_EG_wm_high}
\end{figure}
\GP{Thus the diminishing behavior of the curves $\langle E^\prime\rangle (m=0)$ and $\langle G^\prime\rangle (m=0)$ for decreasing $\omega$ represents a smooth convergence of the moduli to the values for $\omega=0$, similarly to the results in Fig.~\ref{fig_dis_fcc_EG_wm_low}. Experimentally, deviations as small as $0.01 k/l_0^2-0.01 k/l_0^2$ should be accessible within rheometer sensitivities.}

The main difference between the small- and large-$m$ regimes is the qualitative change in $\langle E'_{\alpha\alpha}(\omega)\rangle$ and $\langle G'_{\alpha\beta}(\omega)\rangle$ ($\alpha,\beta=x,y,z$) for increasing magnetic moment.
For $m \lesssim 0.06 m_0$, and according to the different geometries, we observed increase or decrease of the elastic moduli with increasing $m$.
Conversely, for $m \gtrsim 0.06 m_0$ we observe all elastic moduli to increase with increasing magnetic interaction for all frequencies and geometries.

The storage Young's modulus for deformations in the $\vm$-direction $\langle E'_{zz}(\omega)\rangle$ shows the most significant increase when compared to $\langle E'_{xx}(\omega)\rangle$ and $\langle E'_{yy}(\omega)\rangle$.
This trend continues at large $\omega$ [see inset of Fig.~\ref{fig_dis_fcc_EG_wm_high} (a)].
In a similar fashion, the large-$\omega$ behavior of the storage modulus $\langle G'_{zy}(\omega)\rangle$ relative to shear deformations of the chains aligned along $\vm$ [see inset of Fig.~\ref{fig_dis_fcc_EG_wm_high} (b)] suggests a larger increase than for $\langle G'_{xz}(\omega)\rangle$ and $\langle G'_{xy}(\omega)\rangle$.
These overall trends of the dynamic moduli are further enhanced and increased for even larger $m$.

The loss components of the dynamic moduli, both Young and shear, show again an increase with increasing $m$ over all frequencies and geometries.
Furthermore, the amount of increase follows \GP{approximately} the same trends as for the corresponding storage components (see appendix \ref{loss_app}, Fig.~\ref{fig_app_loss_dis_fcc_high}).

\section{Conclusions}\label{conclusions}
We have described and applied a method to determine the dynamic elastic moduli in discretized mesoscopic model systems representing magnetic elastic composite materials.
More precisely, we have confined ourselves to particle-based dipole-spring models \cite{annunziata2013hardening, pessot2014structural, tarama2014tunable, sanchez2013effects,cerda2013phase,ivaneyko2015dynamic} to characterize the behavior of magnetic gels and elastomers.
The magnitudes of Young and shear moduli were evaluated for different frequencies, particle distributions, magnitudes and orientations of the magnetic moments.
We find the elastic moduli to lowest order to increase or decrease with the magnitude of the magnetic moment according to the particle distribution, the selected orientation, and the selected frequency.

To summarize our results, we find that increasing magnetic interactions tend to line up the particles in the direction of the magnetic dipoles.
This, in regular lattices, can result in different effects according to the considered structure.
In general, however, we find the Young modulus in the directions of elongation to increase \cite{ivaneyko2012effects} and, vice versa, to decrease in the directions of shrinking.
For randomized particle arrangements we find a ``hardened'' regime, where dipole--dipole attractions overcome the elastic spring interactions and the elastic moduli significantly increase.
Here, the increase of the storage part of the Young modulus in the direction parallel to the magnetic moments is significantly larger compared to the perpendicular directions, in agreement with experiments reported in the literature \cite{varga2006magnetic,schubert2016equi}.
Furthermore, for all distributions (except for the randomized arrangements at high $m$) we find the storage part of some of the investigated shear moduli to change tendency from increase to decrease with $m$, or vice versa, for increasing frequency $\omega$.
The loss component of the dynamic moduli follows an overall linear behavior for all cases at low and high $\omega$ with a crossover regime in between.
In conclusion, the behavior of the dynamic elastic moduli with varying $m$ and $\omega$ strongly depends on the spatial arrangement of the magnetic particles.
The angles between the magnetic moments and the directions to find the nearest neighbors are crucial to determine whether, for a selected direction, the system shrinks or elongates when switching on magnetic interactions and whether the elastic moduli increase or decrease.

Our systems were of cubelike shape and finite size.
On two opposing boundaries, we imposed prescribed force fields leading to an overall strain response of the whole system.
The other boundaries remained unconstrained.
Such a geometry is characteristic for experimental investigations using plate--plate rheometers.
Assuming particle sizes in the micrometer range, our systems correspond to samples of several ten micrometers in thickness.
Such experimental samples can be analyzed using piezorheometric devices \cite{zanna2002influence,roth2010viscoelastic}.
In fact, for anisotropic magnetic gels, corresponding piezorheometric \GP{measurements} were performed already more than a decade ago \cite{collin2003frozen}.
It will be interesting to compare our approach in more detail with such experimental investigations in the future.

It is important to model and understand the dynamic response of the materials at different frequencies in the view of many practical applications, from soft actuators \cite{zimmermann2006modelling} to vibration absorbers \cite{deng2006development,sun2008study}.
Our method explicitly connects the relaxational modes of the system on the mesoscopic level \cite{tarama2014tunable} with the macroscopic dynamic response \cite{brand2014macroscopic,bohlius2004macroscopic,brand2015macroscopic,menzel2016hydrodynamic}.
Our approach allows to capture the internal rearrangements of the system under an externally applied stress or magnetic field and to link it to the consequences for the overall system behavior.
Furthermore, our technique can be applied to any particle distribution, particularly also to those drawn from experimental analysis of real samples \cite{pessot2014structural,tarama2014tunable}.

Generalizations to systems composed of anisotropic particles \cite{passow2015depolarized}, as well as including rotational degrees of freedom \cite{annunziata2013hardening,weeber2015ferrogels2} and possibly induced-dipole effects \cite{biller2014modeling,allahyarov2015simulation} could be added to the present framework in subsequent steps.
Apart from that, the mesoscopically based dynamic investigations could be extended to more refined approaches, where the elastic matrix between discretized particles of finite volume is described in terms of continuum elasticity theory \cite{cremer2015tailoring}.
As indicated above, it will be possible to use experimental data \cite{gunther2016method,gundermann2014investigation,pessot2014structural,tarama2014tunable} as input for the initial particle positions and compare calculated dynamic moduli with their measured counterparts, also as a function of magnetic interaction strengths.
In a combined effort between experiments and theory, such an approach can serve to devise smarter and new materials with optimized magnetic field dependence and adjusted behavior at different frequencies.

\begin{acknowledgments}
The authors thank the Deutsche Forschungsgemeinschaft for support of this work through the priority program SPP 1681.
\end{acknowledgments}

\newpage
\appendix
\section{Steric Repulsion Parameters}\label{steric_param_app}
The relatively soft steric repulsion between two particles $i$ and $j$ at positions $\bm{R}_i$ and $\bm{R}_j$ connected by the vector $\vr_{ij}= \bm{R}_j -\bm{R}_i$ is modeled by a generic potential $v^{s}(r_{ij})$.
Introducing the exponents $p$ and $q$, the functional form of this potential is given by
\begin{align}\label{generic_strep}
v^{s}(r)&=\varepsilon^{s} \Biggl[ \left( {\frac{r}{\sigma^{s}}} \right)^{-p}- \left( {\frac{r}{\sigma^{s}}} \right)^{-q} -\left( {\frac{r_c}{\sigma^{s}}} \right)^{-p} +\left( {\frac{r_c}{\sigma^{s}}} \right)^{-q} \nonumber \\
& \qquad \qquad -\frac{c^{s}{(r-r_c)}^2}{2} \Biggr]
\end{align}
if $r=|\vr_{ij}|<r_c$ and $v^{s}(r)=0$ otherwise.
The parameter $r_c=\sigma^s(p/q)^{1/(p-q)}$ follows from the condition $v^{s} (r_c)'=0$, whereas $c^s$ is chosen such that $v^{s} (r_c)''=0$.
We find
\begin{equation}
 c^s = \frac{p^{-\frac{2+q}{p-q}}(p-q)q^{\frac{2+p}{p-q}}}{\left({\sigma^s}\right)^2}.
\end{equation}

\section{Derivatives of Pair Interaction Potentials}\label{der_app}
We consider pair interactions between particles $i$ and $j$, at positions $\bm{R}_i$ and $\bm{R}_j$, respectively, and connected by $\vr_{ij}=\bm{R}_j-\bm{R}_i$.
When the particles are linked by a harmonic spring, their harmonic pair interaction potential is
\begin{equation}
 v^{el}_{ij} = \frac{k}{2\ell_{ij}^0} {\left( r_{ij} -\ell_{ij}^0 \right)}^2,
\end{equation}
compare with Eq.~(\ref{eel}).
$r_{ij}=|\vr_{ij}|$ and $l_{ij}^0$ is the unstrained length of the spring.
The gradient components ($\alpha=x,y,z$) follow as (we here drop the $_{ij}$ subscripts for simplicity)
\begin{equation}
 \frac{\partial v^{el}}{\partial r^\alpha} = \frac{k}{\ell^0} \left( r -\ell^0 \right) \ \frac{r^\alpha}{r}.
\end{equation}
The derivatives appearing below in Eq.~(\ref{hessian_pair_pot}) are then
\begin{align}\label{elint_sder}
\mpder{v^{el}}{r^\beta}{r^\alpha} = \frac{k}{\ell^0} \left[ \frac{r^\alpha r^\beta}{r^2}
    + (r - \ell^0) \ \frac{\delta^{\alpha\beta}r^2 - r^\alpha r^\beta}{r^3} \right].
\end{align}

Furthermore, the steric repulsion pair potential $v^{s}$ has been addressed in detail in Appendix \ref{steric_param_app}.
The gradient components ($\alpha=x,y,z$) of the steric pair potential [see Eq.~(\ref{generic_strep})] follow for $r<r_c$ as
\begin{equation}
 \frac{\partial v^{s}}{\partial r^\alpha} =  \frac{-\varepsilon^{s} r^\alpha}{r} \left[{ \frac{p}{r}{\left(\frac{r}{\sigma^s}\right)}^{-p} -\frac{q}{r}{\left(\frac{r}{\sigma^s}\right)}^{-q} +c^s\left( r-r_c \right) }\right]
\end{equation}
and vanish for $r\geq r_c$.
The derivatives below contributing to Eq.~(\ref{hessian_pair_pot}) are given by
\begin{align}\label{strep_sder}
\mpder{v^{s}}{r^\beta}{r^\alpha} = &-\varepsilon^{s} \biggl\{ \left( \frac{\delta^{\alpha \beta}}{r^2}  -2 \frac{r^\alpha r^\beta}{r^4} \right) \left[ p{\left(\frac{r}{\sigma^s}\right)}^{-p} -q{\left(\frac{r}{\sigma^s}\right)}^{-q} \right] \nonumber \\
&-\frac{r^\alpha r^\beta}{r^4}\left[ p^2{\left(\frac{r}{\sigma^s}\right)}^{-p} -q^2{\left(\frac{r}{\sigma^s}\right)}^{-q} \right] \nonumber \\
& +c^s\left[ \frac{r^\alpha r^\beta}{r^2} + (r-r_c)\frac{\delta^{\alpha \beta}r^2 -r^\alpha r^\beta}{r^3} \right] \biggr\}
\end{align}
for $r<r_c$ and vanish when $r\geq r_c$.

Finally, the magnetic pair interaction potential $v^{m}$ as in Eq.~(\ref{emagn}) reads
\begin{align}
 v^{m}_{ij} &=  \frac{\ m^2 r_{ij}^2 -3{(\vm\cdot \rij)}^2}{r_{ij}^5}
\end{align}
in using reduced units, see also Eq.~(\ref{emagn}).
The gradient components ($\alpha=x,y,z$) of the previous expression read
\begin{align}
 \pder{v^{m}}{r^\alpha} = -\frac{3}{r^5} &\Bigl[ m^2 r^\alpha +2m^\alpha(\vm \cdot \vr) \nonumber \\
 & -5 \frac{r^\alpha {(\vm \cdot \vr)}^2}{r^2} \Bigr].
\end{align}
The derivatives appearing below in Eq.~(\ref{hessian_pair_pot}) are given by
\begin{align}\label{dipdip_sder}
& \mpder{v^{m}}{r^\beta}{r^\alpha} = -\frac{3}{r^5}\ \Biggl[ m^2 \delta^{\alpha \beta} -5m^2 r^\alpha r^\beta r^{-2} \nonumber \\
& -10({\vm}\cdot\vr)r^{-2}\left(m^\alpha r^\beta+m^\beta r^\alpha\right) +2m^\alpha m^\beta \nonumber \\
& -5\left(\vm\cdot\vr\right)^2r^{-2}\left(\delta^{\alpha\beta}-7 r^\alpha r^\beta r^{-2}\right)\Biggr].
\end{align}

\section{Hessian Matrix for Pair Interaction Potentials}\label{hess_app}
Here we repeat in detail the derivation of the Hessian for a system interacting solely via pair potentials.
That is, any two particles $i$ and $j$ at positions $\bm{R}_i$ and $\bm{R}_j$ interact through a pair potential $v$ depending only on the connecting vector $\vr_{ij}= \bm{R}_j -\bm{R}_i$.
Then we can write
\begin{equation}\label{eqapp1}
 U = \frac{1}{2} \sumtwolines{i,j=1}{i\neq j}{N} v(\vr_{ij}),
\end{equation}
where $N$ is the total number of particles.
Again, $\bm{R}_i$ is the position of the $i$-th particle ($i=1\dots N$), $\vr_{ij} = \bm{R}_j -\bm{R}_i$, and we denote by $R_i^\alpha$ ($\alpha=x,y,z$) the $\alpha$-component of $\bm{R}_i$.
For reasons of symmetry, $v(\vr_{ij})= v(\vr_{ji})$.
The sum in Eq.~(\ref{eqapp1}) together with the prefactor $\frac12$ then runs over all different pairs counting each of them only once.
We abbreviate $v_{ij}=v(\vr_{ij})$.
The gradient components ($\alpha=x,y,z$) of the energy $U$ follow as
\begin{align}
 \pder{U}{R_k^\alpha} &=\frac{1}{2}\sumtwolines{i,j=1}{i\neq j}{N} \pder{v_{ij}}{R_k^\alpha} \\
 &= \sumtwolines{j=1}{j\neq k}{N}\pder{v_{kj}}{R_k^\alpha}= -\sumtwolines{j=1}{j\neq k}{N} \pder{v_{kj}}{r_{kj}^\alpha}, \nonumber
\end{align}
setting the force $-\partial{U}/\partial{\vR_k}$ on the positional degrees of freedom of the $k$-th particle.

Next, we obtain the Hessian of the system as
\begin{equation}
\mpder{U}{R_i^\alpha}{R_k^\beta}=\left\{ {
\begin{array}{*{20}cl}
   \mpder{v_{ik}}{R_i^\alpha}{R_k^\beta}  &(i\neq k),  \\
   \\
   \displaystyle \sumtwolines{j=1}{j\neq i}{N} \mpder{v_{ij}}{R_i^\alpha}{R_i^\beta}  & (i= k).\\
\end{array}
} \right.
\end{equation}
Thus, for pair interactions, the diagonal elements of the Hessian contain the second derivatives of all pair interactions, whereas the off-diagonal elements are given by a single term.
Since $\vr_{ij}=\vR_j-\vR_i$, the previous equation can be expressed in terms of connecting vectors only:
\begin{equation}\label{hessian_pair_pot}
\mpder{U}{R_i^\alpha}{R_k^\beta}=\left\{ {
\begin{array}{*{20}cl}
   -\mpder{v_{ik}}{r_{ik}^\alpha}{r_{ik}^\beta} &(i\neq k), \\
   \\
   \displaystyle \sumtwolines{j=1}{j\neq i}{N} \mpder{v_{ij}}{r_{ij}^\alpha}{r_{ij}^\beta} &(i= k).\\
\end{array}
} \right.
\end{equation}

\section{Torque-Free Force Fields}\label{torq_app}
Our scope is to describe the system behavior for pre\-selected specified orientations.
However, both during the search for the corresponding equilibrium state of the system (see section \ref{equil_state}) and the implementation of an external force (see section \ref{micr_stress}), the system may tend to perform a rigid rotation.
We therefore must exclude such rigid rotations.
Here we describe a simple method to redefine the generalized force field (or likewise the gradient of the total energy) so that the net overall torque on the system vanishes.

We consider the force field $\bm{f}$ acting on the particles at positions $\vR_i$ with components $\bm{f}_i$ ($i=1,\dots N$).
The net torque $\bm{\tau}$ is given by
\begin{equation}
 \bm{\tau} = \sum_{i=1}^N \vq_i \times \bm{f}_i,
\end{equation}
where $\vq_i = \vR_i - \vR_c$ is the distance of the particle positions $\vR_i$ from the center of mass $\vR_c =\frac1N \sum_i \vR_i$.
To prevent, e.g.,\ a global rotation of the system around the $z$-axis, the $z$-component of $\bm{\tau}$, i.e.\ $\tau^z$, must vanish.
We define a uniform, counter-clockwise rotational force field around the $z$-axis $\bm{P}\left(\vq\right) = c_R (-q^y, q^x, 0)$, with $\vq$ a vector in the $xy$-plane and $c_R$ a constant.
Next, we determine $c_R$ by imposing $\bm{P}$ to have the same torque as given by $\bm{f}$:
\begin{align}
 \sum_{i=1}^N {\left( \vq_i \times \bm{f}_i \right)}^z = \tau^z &= \sum_{i=1}^N {\left[ \vq_i \times \bm{P}\left(\vq_{i}\right) \right]}^z \\ 
& = c_R \sum_{i=1}^N \left[ {\left(q_i^x\right)}^2 + {\left(q_i^y\right)}^2 \right]. \nonumber
\end{align}
We obtain the field $\bm{P}$ by solving for the constant $c_R$, leading to
\begin{equation}
 c_R = \frac{\tau^z}{\sum_{i=1}^N \left[ {\left(q_i^x\right)}^2 + {\left(q_i^y\right)}^2\right]}.
\end{equation}
Therefore we can make $\bm{f}$ ``torque-free'' concerning the $z$-direction by subtracting $\bm{P}$, i.e.\ $\bm{f}_i \rightarrow \bm{f}_i -\bm{P}(\vq_i)$ ($i=1,\dots N$).
By repeating the procedure for the remaining directions, we get rid of the rigid rotations induced by $\bm{f}$ and obtain a torque-free force field.

\section{Static Young Moduli of Regular Lattices}\label{anyoung_app}
We here present a simple energy argument to interpret the behavior of the Young moduli of the regular lattices presented in sections \ref{cubic_result} and \ref{fcc_result}.
A regular lattice is generated by the basis vectors $\bm{a}_1$, $\bm{a}_2$, and $\bm{a}_3$.
Therefore a lattice point can be written as $\bm{r}_{ijk}=i\bm{a}_1+j\bm{a}_2+k\bm{a}_3$, with $i,j,k \in \mathbb{Z}$ integers.
If the particles interact by the pair potential $v$, the total energy per particle in an infinitely extended lattice is given by
\begin{equation}\label{latt_en}
 U_p=\frac{1}{2}\sum_{n\in \mathcal{N}_0} v(\bm{r}_n),
\end{equation}
where the sum runs over all lattice points (origin excluded) labeled by the discrete index $n$ contained in the set $\mathcal{N}_0=\mathbb{Z}^3\setminus\{(0,0,0)\}$.

Since we consider the regular lattice to be the ground state of the system, a small deformation that transforms $\bm{r}_n\rightarrow\bm{r}'_n$ ($n\in\mathcal{N}_0$) has an energy-per-particle cost that to lowest order reads
\begin{equation}
 \Delta U_p=\frac{1}{2}\sum_{n\in \mathcal{N}_0} \frac{1}{2} \bm{u}_n^\intercal\cdot \mathbf{h}(\bm{r}_n) \cdot\bm{u}_n
\end{equation}
where $\bm{u}_n=\bm{r}'_n-\bm{r}_n$, $^\intercal$ indicates transposition, and $\mathbf{h}(\bm{r}_n)$ is the Hessian matrix of the interaction $v(\bm{r}_n)$ between the particle fixed in the origin and the $n$th neighbor.
Its elements are given by $\textrm{h}^{\mu\nu}(\bm{r}_n)=\partial^2v(\bm{r}_n)/\partial {r}_n^\mu\partial {r}_n^\nu$, with $\mu,\nu=x,y,z$.

The displacements $\bm{u}_n = \mathbf{D}\cdot\bm{r}_n$ corresponding to a uniform strain are given by the constant components of the displacement tensor $\mathbf{D}$.
The energy of the strain deformation then follows as
\begin{align}\label{an_stiff_mod}
 \Delta U_p&=\frac12 \sum_{\alpha\beta\gamma\delta} C_0^{\alpha\beta\gamma\delta}{D}^{\alpha\beta}{D}^{\gamma\delta}\nonumber \\
 \mbox{with}\ C_0^{\alpha\beta\gamma\delta}&=\frac12 \sum_{n\in \mathcal{N}_0} {r}_n^\alpha\ \textrm{h}^{\beta\gamma}(\bm{r}_n)\ {r}_n^\delta,
\end{align}
where $\alpha,\beta,\gamma,\delta=x,y,z$.

In the following we focus on compressive/dilative strains and therefore consider diagonal $\mathbf{D}$ displacement tensors.
For an applied strain $\varepsilon_{\alpha\alpha}$ along the $\alpha$-direction $\textrm{D}^{\alpha\alpha}\neq 0$ is imposed.
The remaining components of $\mathbf{D}$ are relaxed to minimize the lattice energy
\begin{equation}
\frac{\partial \Delta U_p}{\partial \textrm{D}^{\mu\mu}}=0,\ \ \ \forall \mu\neq\alpha.
\end{equation}
This leads to a system of linear equations the solution of which relates the components $\textrm{D}^{\mu\mu}$ ($\mu\neq\alpha$) to the imposed deformation $\textrm{D}^{\alpha\alpha}$.
As a result, we obtain Young's modulus $E_{\alpha\alpha}$ [following the notation as in the main text, see Eq.~(\ref{macr_elm})] given by
\begin{align}
 E_{\alpha\alpha}&=\frac{1}{V_p}\frac{\d^2 \Delta U_p}{{\left(\d \textrm{D}^{\alpha\alpha}\right)}^2}=\frac{1}{V_p}\left(C_0^{\alpha\alpha}-B^{\alpha}\right) \nonumber \\
\mbox{with}\ B^{\alpha}&= \sum_{\beta\gamma} C_0^{\alpha\beta}\frac{C_0^{\gamma\gamma}C_0^{\alpha\beta}-C_0^{\alpha\gamma}C_0^{\beta\gamma}}{C_0^{\beta\beta}C_0^{\gamma\gamma}-{ (C_0^{\beta\gamma}) }^2}{(\epsilon^{\alpha\beta\gamma})}^2,
\end{align}
where $V_p=1/\rho=V/N$ is the volume per particle, we abbreviated $C_0^{\alpha\beta}=C_0^{\alpha\alpha\beta\beta}$, and $\epsilon^{\alpha\beta\gamma}$ is the Levi-Civita symbol.
The contributions $-B^{\alpha}$ to the elastic moduli take into account relaxation along the remaining perpendicular axes and lower the moduli.

For small values of the magnetic moment $m$, we write, to lowest order in $m$, $\mathbf{h}(\bm{r}_n)=\mathbf{h}_0(\bm{r}_n)+m^2 \mathbf{h}_{m}(\bm{r}_n)$, where the elements of the matrix $m^2 \mathbf{h}_{m}(\bm{r}_n)$ are as listed in Eq.~(\ref{dipdip_sder}).
Thus, we can obtain both the static Young's modulus at $m=0$ and the initial quadratic behavior for small $m$.

\section{Additional Information on the Loss Part of the Dynamic Elastic Moduli}\label{loss_app}
Here we show in more detail the various behaviors of the loss part of the dynamic moduli as functions of frequency $\omega$ and magnitude of the magnetic moment $m$ for the different considered geometries.
As we have mentioned before, we find as a general trend the loss parts to linearly increase with $\omega$ at low and high frequencies.
It results from our viscous friction term [see Eq.~(\ref{eq_motion})] which, in the case of an oscillatory deformation as in Eq.~(\ref{osc_displ}), is proportional to $\omega$.
Moreover, it is consistent with the predicted loss component of the dynamic moduli in the Kelvin-Voigt model \cite{mainardi2011creep,eldred1995kelvin}.
Therefore, and for better illustration, we plot the loss parts after division by $\omega$.
The agreement with linear behavior is confirmed in this way, i.e.\ $E''_{\alpha\alpha}(\omega)/\omega$ and $G''_{\alpha\beta}(\omega)/\omega$ ($\alpha,\beta=x,y,z$) converge to a finite value in both the low- and high-$\omega$ limit, see Figs.~\ref{fig_app_loss_cubic}--\ref{fig_app_loss_dis_fcc_high}.
\begin{figure}[]
\centering
  \includegraphics[width=8.6cm]{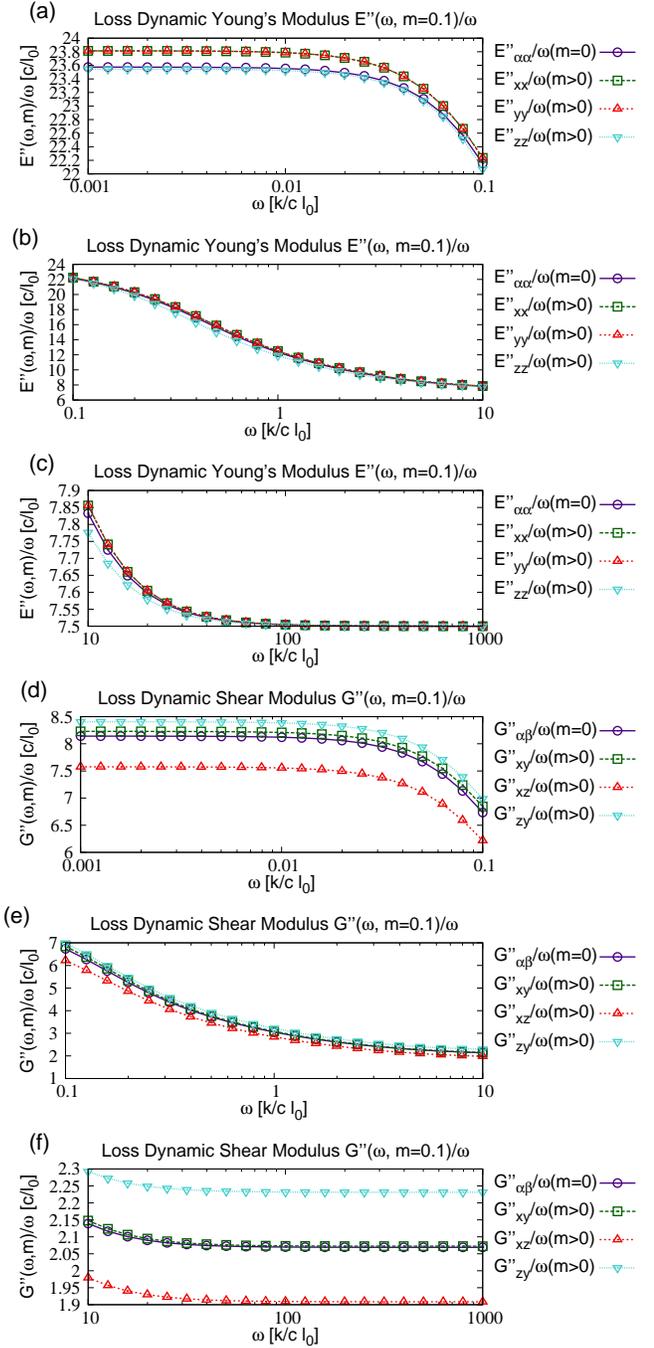}
  \caption{Loss parts (a, b, c) $E''_{\alpha\alpha}(\omega)$ and (d, e, f) $G''_{\alpha\beta}(\omega)$ ($\alpha,\beta=x,y,z$) for the dynamic elastic moduli of a simple cubic lattice with $N=\npcube$ for vanishing magnetic moment (solid line, $\bigcirc$), and $\vm=0.1 m_0 \dirz$ (dashed lines, $\Box,\bigtriangleup,\bigtriangledown$).
Because of the overall trend of a linear increase in frequency at low and high frequencies, we here present the moduli divided by $\omega$.
Zoom-ins onto the low-, intermediate-, and high-$\omega$ regions are shown in panels (a, d), (b, e) and (c, f), respectively.
}
  \label{fig_app_loss_cubic}
\end{figure}
\begin{figure}[]
\centering
  \includegraphics[width=8.6cm]{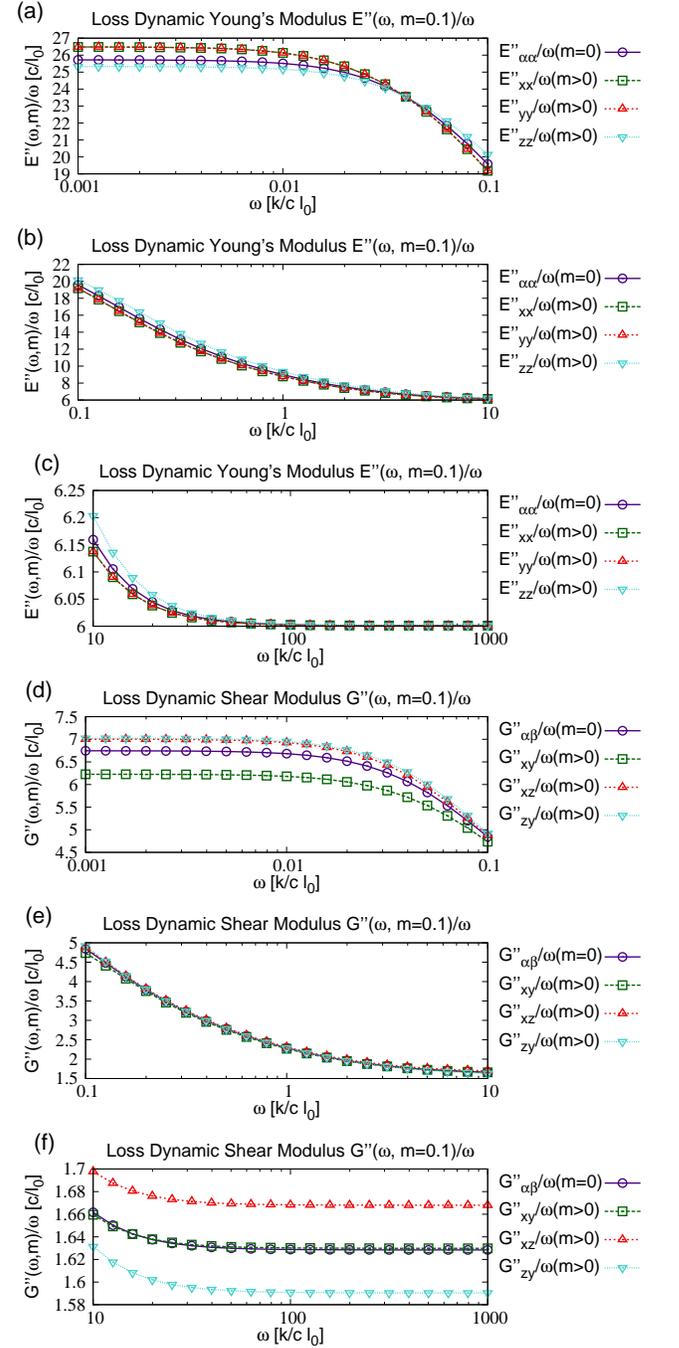}
  \caption{Loss parts (a, b, c) $E''_{\alpha\alpha}(\omega)$ and (d, e, f) $G''_{\alpha\beta}(\omega)$ ($\alpha,\beta=x,y,z$) for the dynamic elastic moduli of an \fcc\ lattice with $N=\npfcc$ for vanishing magnetic moment (solid line, $\bigcirc$), and $\vm=0.1 m_0 \dirz$ (dashed lines, $\Box,\bigtriangleup,\bigtriangledown$).
Since the loss moduli increase linearly with the frequency at low and high frequencies, we here show them divided by $\omega$.
Zoom-ins onto the low-, intermediate-, and high-$\omega$ regions are shown in panels (a, d), (b, e) and (c, f), respectively.
}
  \label{fig_app_loss_fcc}
\end{figure}
\begin{figure}[]
\centering
  \includegraphics[width=8.6cm]{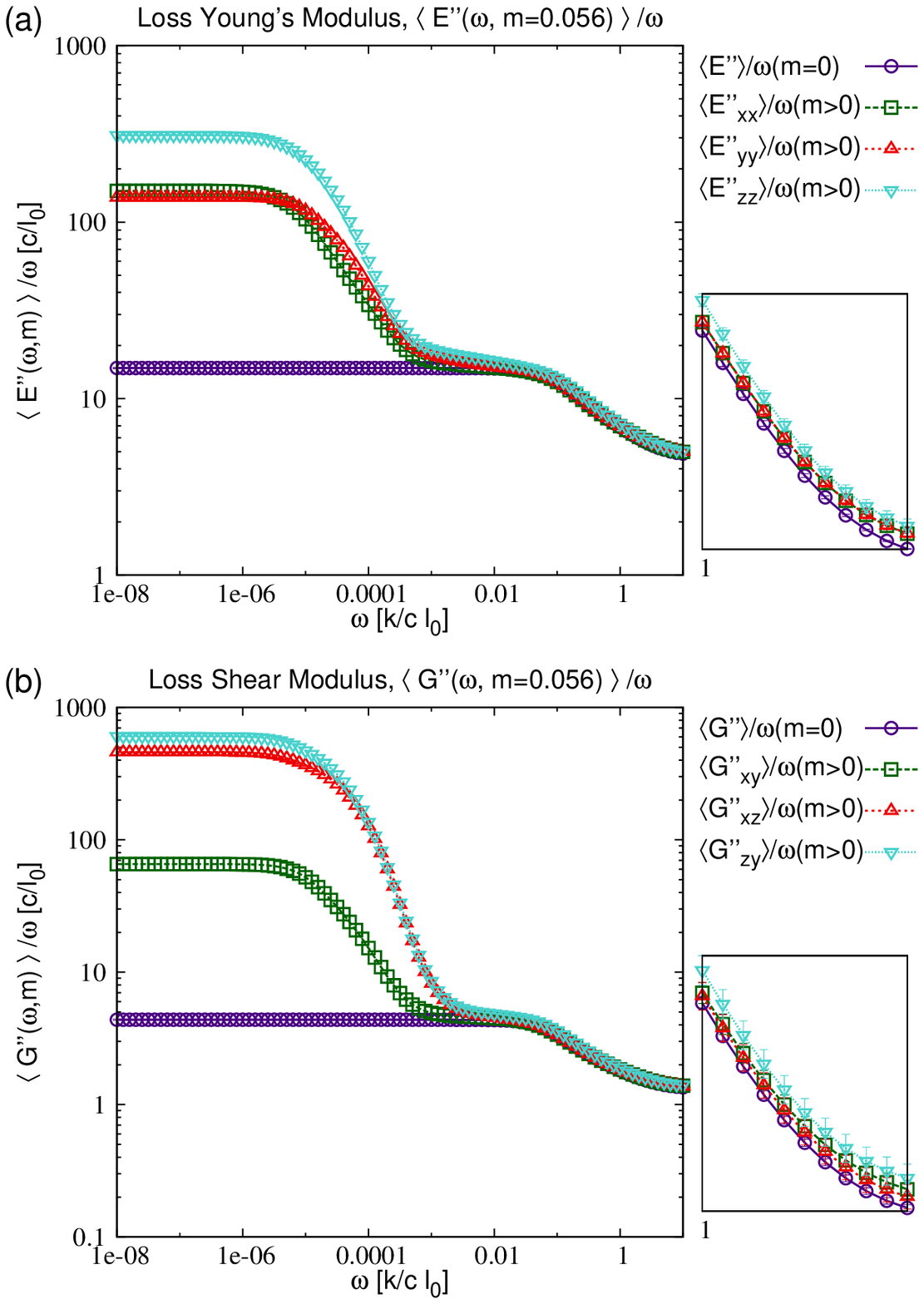}
  \caption{Average loss parts (a, b) $\langle E''_{\alpha\alpha}(\omega)\rangle$ and (c, d) $\langle G''_{\alpha\beta}(\omega)\rangle$ ($\alpha,\beta=x,y,z$) for the dynamic elastic moduli of randomized \fcc\ lattices with $N=\npdisord$ for vanishing magnetic moment (solid line, $\bigcirc$), and $\vm=\mlow m_0 \dirz$ (dashed lines, $\Box,\bigtriangleup,\bigtriangledown$).
Because of the overall trend of a linear increase in frequency at low and high frequencies, we here present the moduli divided by $\omega$.
Data points and \GP{standard deviations} are obtained by averaging over $\ndissamples$ differently randomized samples.
Because of the different randomizations, the initial slope of the moduli in the $\omega\rightarrow 0$ limit can vary significantly, thus leading to large bars in the small-$\omega$ regime and for the $m>0$ cases, which are not shown here.
Insets (a) and (b) zoom in onto the Young and shear loss moduli behavior, respectively, at high frequencies for better resolving the individual curves.
}
  \label{fig_app_loss_dis_fcc_low}
\end{figure}
\begin{figure}[]
\centering
  \includegraphics[width=8.6cm]{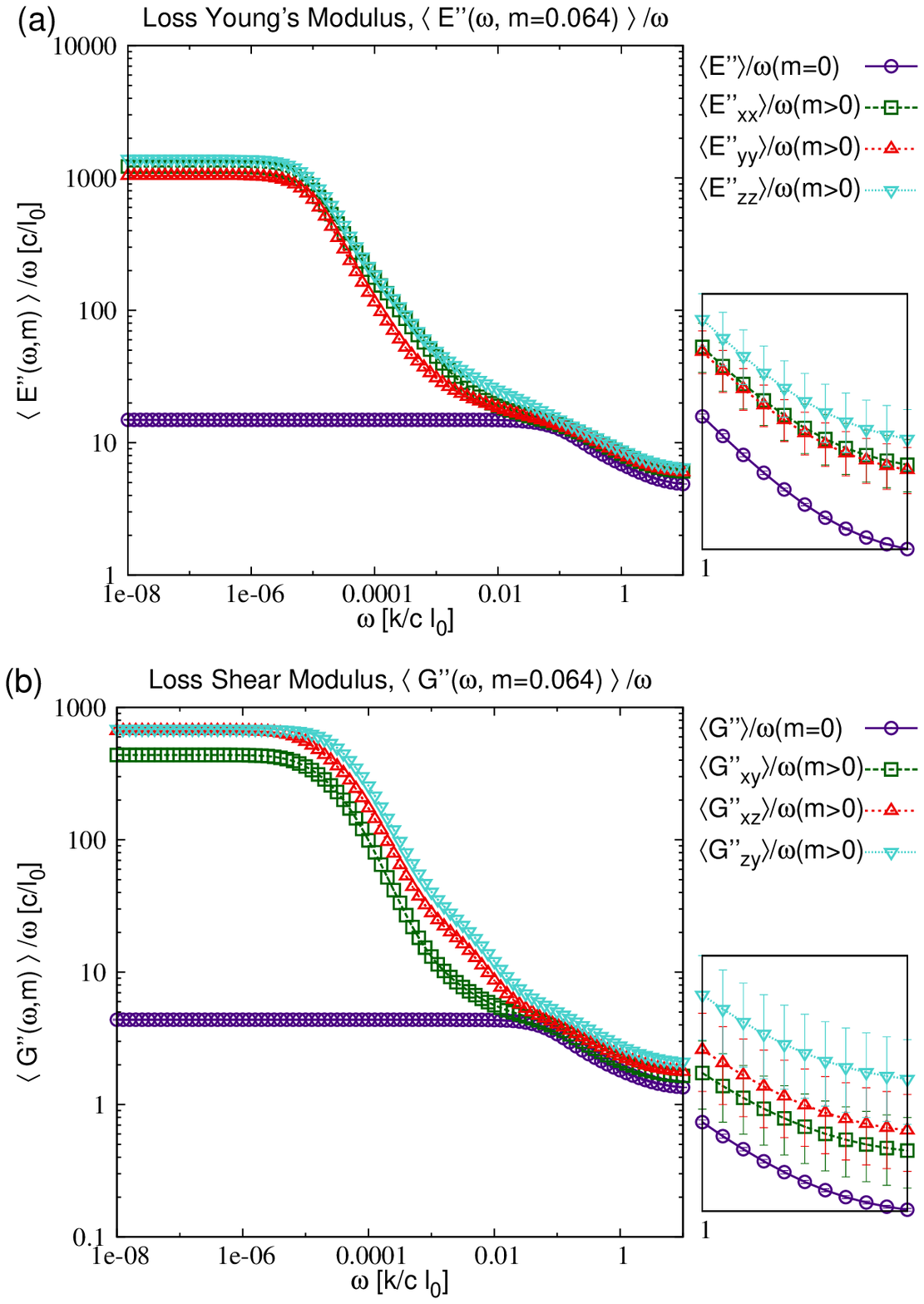}
  \caption{Average loss parts (a, b) $\langle E''_{\alpha\alpha}(\omega)\rangle$ and (c, d) $\langle G''_{\alpha\beta}(\omega)\rangle$ ($\alpha,\beta=x,y,z$) for the dynamic elastic moduli of randomized \fcc\ lattices with $N=\npdisord$ for vanishing magnetic moment (solid line, $\bigcirc$), and $\vm=\mhigh m_0 \dirz$ (dashed lines, $\Box,\bigtriangleup,\bigtriangledown$).
Since the loss moduli increase linearly with the frequency at low and high frequencies, we here show them divided by $\omega$.
Data points and \GP{standard deviations} are obtained by averaging over $\ndissamples$ differently randomized samples.
Because of the different randomizations, the initial slope of the moduli in the $\omega\rightarrow 0$ limit can vary significantly, thus leading to large bars in the small-$\omega$ regime and for the $m>0$ cases, which are not shown here.
Insets zoom in onto the (a) Young and (b) shear loss moduli behavior at high frequencies for better resolving the individual curves.
}
  \label{fig_app_loss_dis_fcc_high}
\end{figure}

On the one hand, the regular lattices addressed in sections \ref{cubic_result} and \ref{fcc_result} show different trends for the loss parts as functions of $m$ and $\omega$, as mentioned in the main text and illustrated in Figs.~\ref{fig_app_loss_cubic} and \ref{fig_app_loss_fcc}.
On the other hand our randomized lattices generally show increasing loss parts with increasing $m$ for all frequencies, although the amount of gain varies with the selected geometries, see Figs.~\ref{fig_app_loss_dis_fcc_low} and \ref{fig_app_loss_dis_fcc_high}.

\textheight=25cm


\end{document}